\documentclass[useAMS,usenatbib]{mn2e}

\usepackage{graphicx}
\usepackage{times}

\begin{document}

\title[The scatter of clusters profiles.] {The scatter in the radial profiles of X-ray luminous galaxy clusters as diagnostic of the thermodynamical state of the ICM}
\author[F. Vazza, M.Roncarelli, S.Ettori, K.Dolag]{F. Vazza$^{1,2}$\thanks{E-mail: f.vazza@jacobs-university.de}, M.Roncarelli$^{3,4}$, S.Ettori$^{5,6}$, K.Dolag$^{7,8}$\\
$^{1}$ Jacobs University Bremen, Campus Ring 1, 28759, Bremen, Germany \\
$^{2}$INAF/Istituto di Radioastronomia, via Gobetti 101, I-40129 Bologna,
Italy\\
$^{3}$ Dipartimento di Astronomia, Universit\`a di Bologna, via Ranzani
 1, I-40127 Bologna, Italy \\
$^{4}$ Centre d'Etude Spatiale des Rayonnements, CNRS/Universit\'e de 
Toulouse, 9 avenue du Colonel Roche, BP44346, 31028 Toulouse 
Cedex 04, France \\
$^{5}$ Istituto Nazionale di Astrofisica --
Osservatorio Astronomico di Bologna, Via Ranzani 1, I-40127 Bologna, Italy\\
$^{6}$ INFN, Sezione di Bologna, viale Berti Pichat 6/2, I-40127 Bologna, Italy\\
$^{7}$  Max-Planck-Institut f\"ur Astrophysik, Garching, Germany\\
$^{8}$ University Observatory Munich, Scheinerstr. 1, D-81679 Munich, Germany}

\date{Received / Accepted}
\maketitle
\begin{abstract}
We study the azimuthal scatter in the radial profiles of X-ray luminous galaxy
clusters, with two sets of high-resolution cosmological re-simulations 
obtained with the codes ENZO and GADGET2. The average gas profiles are computed for different angular sectors of the cluster projected volume, and compared with the mean cluster profiles at each radius from the center. We report that in general the level of 
azimuthal scatter is found to be $\sim 10$ per cent for gas density,
temperature and entropy inside $R_{200}$, and $\sim 25$ per cent for X-ray luminosity for the same volume. These
values generally doubles going to $2 R_{200}$ from the cluster center, and are generally found to be higher (by  $\sim 20-40$ per cent) in the case of perturbed systems.
 A comparison with
results from recent {\it Suzaku} observations is discussed, showing the 
possibility to simply interpret the large azimuthal scatter of observables in light of our
simulated results. 
\end{abstract}

\label{firstpage} 

\vskip 0.4cm \fontsize{11pt}{11pt} \selectfont

\section{Introduction}
\label{sec:intro}

Only a fraction of the volume occupied by the Intra-Cluster Medium is currently 
mapped with X-ray observations. 
The typical measures of the surface brightness, temperature and
metal abundance extend out to a fraction of the virial radius, $R_{vir}$, and only
in a few cases meaningful estimates at $R_{500}$ {\footnote {
$R_{500}$ ($R_{200}$) is the radius enclosing a mean matter over-density of 500 (200) times the critical density of the Universe. 
 Compared to the cluster virial radius these radii correspond to $R_{\rm 500} \approx 0.5 R_{\rm vir}$ and 
to $R_{\rm 200} \approx 0.7 R_{\rm vir}$.}} or beyond are obtained.
Because of this, more than two-thirds of the typical cluster volume,
where the primordial gas is accreted and the dark matter (DM) halo is
forming, are still unknown for what concerns the gas/DM mass distribution
and the related thermodynamical properties.
This poses a limitation in our ability to characterize 
the physical processes presiding over the formation and evolution of clusters,  
and  thus to use them as high-precision cosmological tools (see Ettori \& Molendi 2010 and references therein).

In general the surface brightness distribution, requiring lower counts statistic,
is easier to be characterized and has been successfully studied with {\it Rosat} PSPC (e.g. the outskirts 
of the Perseus cluster in Ettori et al. 1998; more systematic studies
have been carried out from Vikhlinin et al.  1999 and Neumann 2005)
and with {\it Chandra} (e.g. Ettori \& Balestra 2009).
On the other hand the measure of the gas temperature is more strongly limited
from the background emission. 
Current exposures with {\it Chandra} (e.g. Vikhlinin et al. 2006) and {\it XMM}
(e.g. Leccardi \& Molendi 2008) enabled the reconstruction of the
azimuthally-averaged temperature profile for few tens of nearby galaxy clusters
out to $\sim R_{500}$. A step forward has been recently obtained 
from the observations of a handful sample of nearby X-ray luminous
objects with the Japanese satellite {\it Suzaku} that, despite the relatively poor PSF and small FOV
of its X-ray imaging spectrometer (XIS), benefits from the modest background associated 
to its low Earth orbit (see results on PKS0745-191, George et al. 2009; A2204, Reiprich et al. 2009;
A1795, Bautz et al. 2009; A1413, Hoshino et al. 2010, A1689, Kawaharada et al.2010). However as consequence of the small FOV (and of the large solid angle
covered from the bright nearby clusters) these observations
provide often discrepant constraints over a few arbitrarily chosen directions.

A number of numerical and semi-analitical works in the last few years aimed at providing reliable models of the thermal properties of galaxy clusters at their
outskirts, from the theoretical point of view (e.g.Loken et al.2002;  Roncarelli et al.2006; Burns, Skillman \& O'Shea 2010; Lapi, Fusco-Femiano \& Cavaliere 2010; Cavaliere, Lapi \& Fusco-Femiano 2010; Vazza et al.2010).
Quite  recently, a detailed analysis of the cluster profiles up to $\approx 2 \cdot R_{200}$ was reported by Burns et al.(2010), using  a sample of clusters simulated with the ENZO code (see Sec.\ref{sec:methods}).

In this article, we aim at characterizing the intrinsic azimuthal scatter (due to the 3--D structure of clusters) that should affect at some level the measurements of the 
thermodynamical properties in the cluster outskirts,  and at providing a guidance in
interpreting the existing and incoming observations. To this end,
we extended a method proposed by Roncarelli et al.(2006) to compute the cluster profiles by avoiding the bias
caused by small collapsed gas clumps, and we measured the level of azimuthal scatter around the average clusters
profiles in a total sample of 29 massive galaxy clusters. 

The paper is organized as follows: in Sec.\ref{sec:methods} we describe the set of
simulations;  in Sec.\ref{sec:results} we present the average properties of
the azimuthal scatter of the gas density, temperature, entropy and X-ray luminosity,
with a comparison with the observational constraints available.
The discussion of our results is reported in Sec.\ref{sec:conclusions}. In the Appendix, additional test on the
physical and numerical convergence of the main results of the paper are discussed.

\begin{figure} 
\includegraphics[width=0.49\textwidth, height=0.4\textwidth]{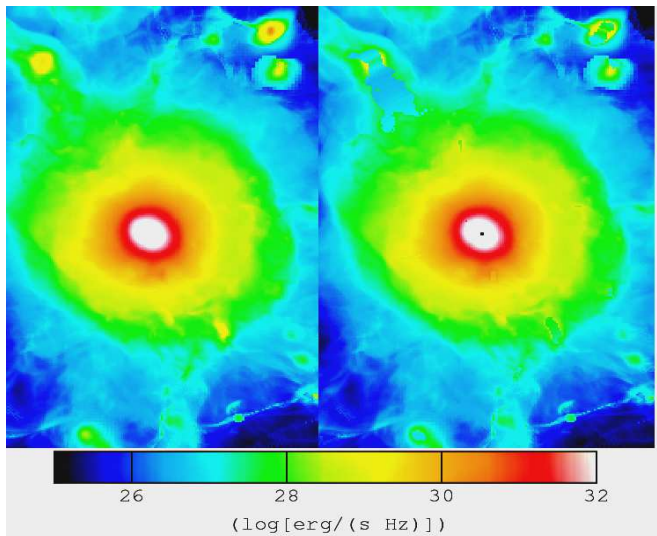}
\caption{Projected X-ray luminosity from each pixel at $[0.5-2]$keV for cluster E14 
of the ENZO sample at $z=0$. The left panel is for the original data, the right panel
is for the data after the 99 per cent filtering. The side of the slices 
is $\sim 3 \times 5$Mpc/h, the pixel size is $25kpc/h$.}
\label{fig:lx_smooth}
\end{figure}

\begin{figure} 
\includegraphics[width=0.45\textwidth,height=0.4\textwidth]{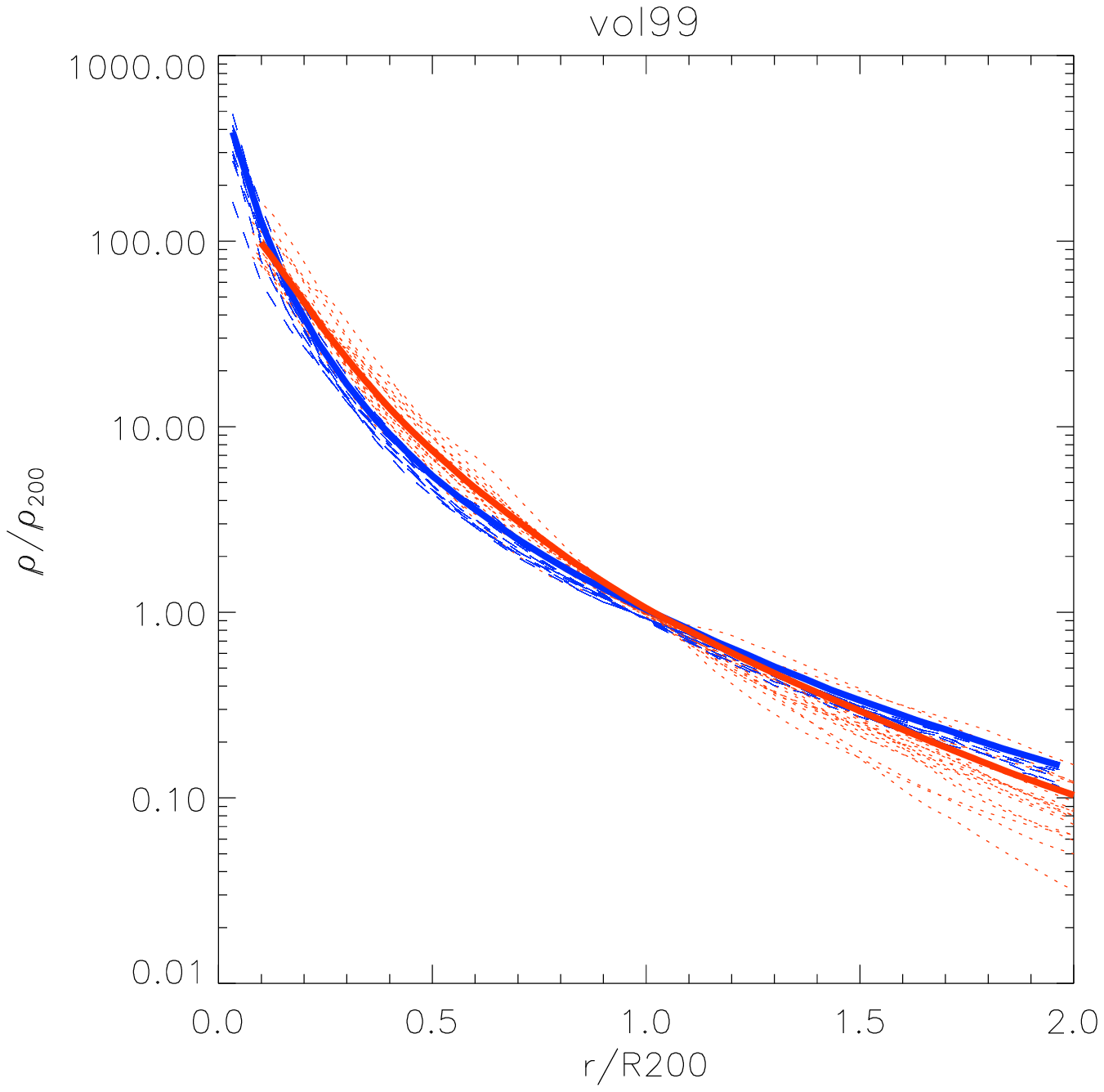}
\includegraphics[width=0.45\textwidth,height=0.4\textwidth]{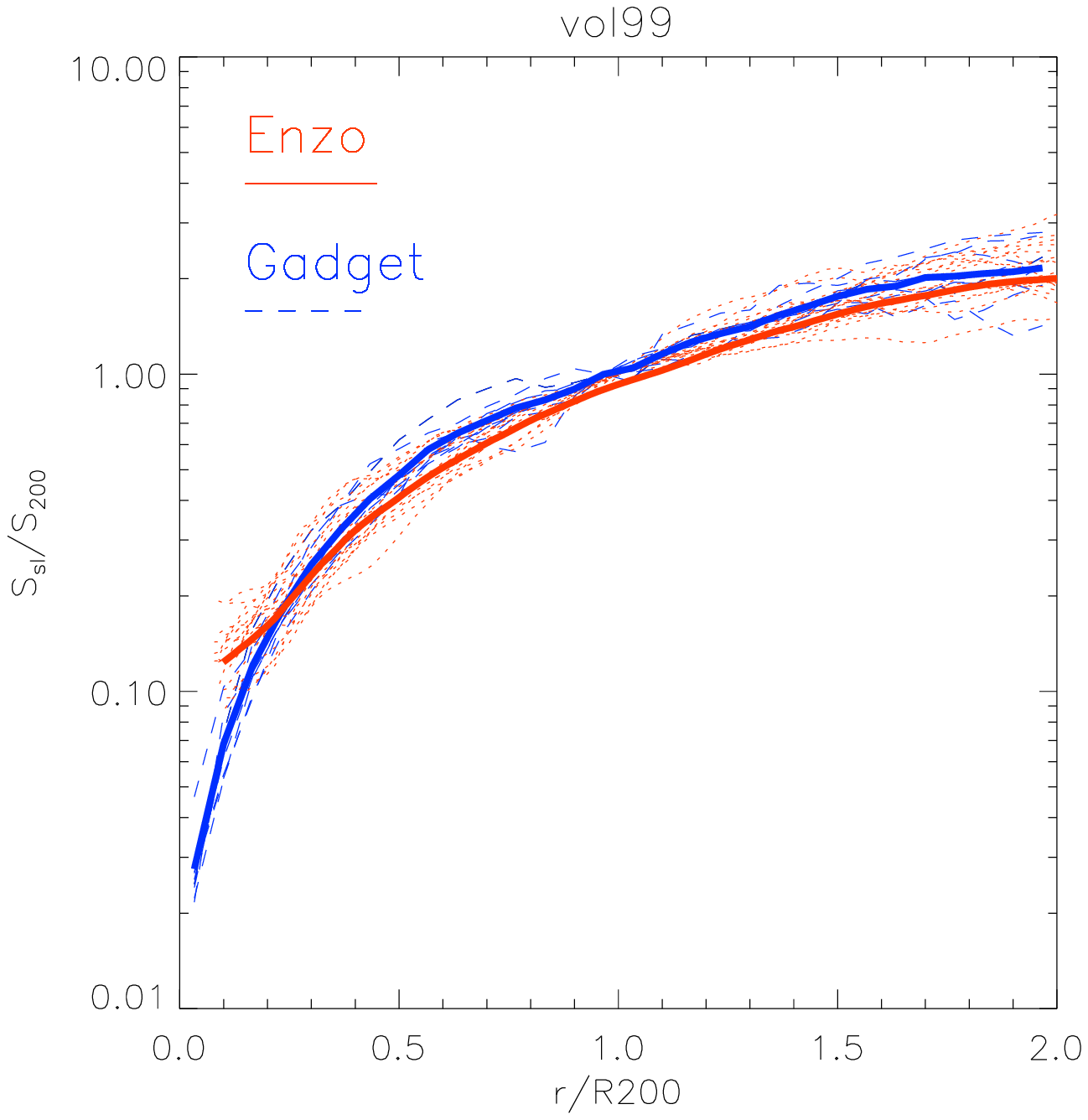}
\includegraphics[width=0.45\textwidth,height=0.4\textwidth]{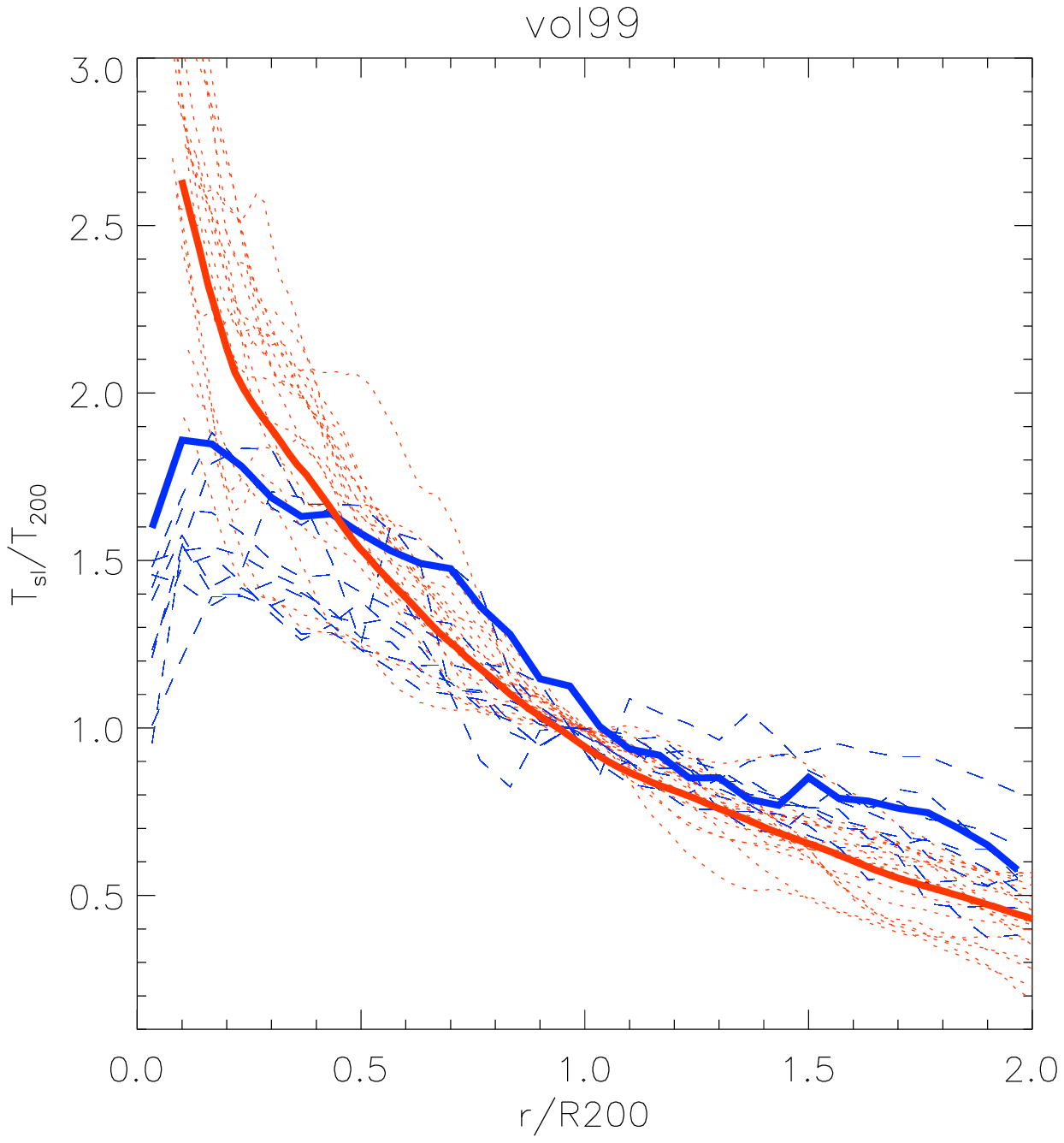}
\caption{Filtered profiles of gas density (top), gas entropy (center) and
gas temperature (bottom) for all GADGET clusters (blue) and for all ENZO clusters (red).
The profiles of all quantities are normalized at $R_{200}$. The dotted/dashed lines refer to each cluster in the sample, the thick solid lines 
show the average profiles for each of the two samples.}
\label{fig:prof_temp}
\end{figure}

\section{Numerical methods}
\label{sec:methods}
\subsection{Cluster simulations}

We analyze two independent sets of high-resolution ($\delta_{\rm max} \sim 10-20$kpc/h) 
cosmological 
simulations of galaxy clusters, extracted from large cosmological boxes
($L_{\rm box} \sim 500$ Mpc) and re-simulated  with two of the most tested numerical codes for 
high performance computing: the Adapative Mesh Refinement code  ENZO and the Tree Smoothed Particles Hydrodynamics code GADGET2. The main cosmological and numerical parameters adopted are listed in Table 1. 

 A  data-set of 20 non-radiative re-simulations of galaxy clusters with total masses 
$6 \cdot 10^{14} \leq M/M_{\odot} \leq 2 \cdot 10^{15}$ was produced with the 1.5 public version of ENZO 1.5{\footnote{ENZO is developed by the Laboratory for Computational
 Astrophysics at the University of California in San Diego 
(http://lca.ucsd.edu)}}(Norman et al.2007 and references therein), with a few physical implementation (Vazza et al.2010, hereafter Va10).
For the specific details about the simulational setup and the post-processing procedure adopted we address the reader to Va10; a
public archive of part of this data-sample is currently available at {\it http://data.cineca.it}.

 Massive halos in ENZO runs were identified following a spherical over-density method, applied
to the simulated outputs at $z=0$ (e.g. Gheller, Pantano \& Moscardini 1998). All clusters with total masses larger than $\sim 6 \cdot 10^{14}M_{\odot}$ were
extracted and sampled at the highest available resolution in the simulation $25kpc/h$ out to the distance of $\sim 3-4R_{200}$ from
the cluster centers.

A second sample of 9 non-radiative re-simulations of galaxy clusters with masses
$2 \cdot 10^{14}\leq M/M_{\odot} \leq 3 \cdot 10^{15}$ was obtained
with GADGET2 (Springel 2005). The details for the simulational setup of these runs can be found in Dolag et al.(2005) 
and Roncarelli et al.(2006), hereafter Ro06.

Also in this case, the halos at $z=0$ were identified basing on a spherical over-density method; the extraction procedure was tailored in order
to have a sample of obviously non-merging systems across the mass range, discarding double systems or systems with ongoing strong interactions.

\begin{table}
\label{tab:tab1}
\caption{Numerical and cosmological parameters for the simulations: code, mass resolution
for DM, gravitational softening, maximum gas spatial resolution $\delta_{\rm max}$, cosmological DM density, cosmological baryon density, normalization of the primordial density power spectrum. GADGET2 runs adopted $h=H/(100km/(s \cdot Mpc))=0.7$ while
ENZO runs adopted $h=0.72$.}
\begin{tabular}{c|c|c|c|c|c|c|c}
code & $M_{dm}$ & $\epsilon$ & $\delta_{\rm max} $ & $\Omega_{DM}$ & $\Omega_{b}$ & $\sigma_{8}$ \\ 
 & $[M_{\odot}/h]$ & $[kpc/h]$ & $[kpc/h] $ &  &  &  \\ 
\hline
{\rm ENZO} & $ 6.7 \cdot 10^{8}$ & 50 & 25 & 0.226 & 0.044 & 0.8 \\
{\rm GADGET2} & $ 1.13 \cdot 10^{9}$ & 5 & $\sim$ 10 & 0.261 & 0.039 & 0.9 \\
 \end{tabular}
\end{table}


\begin{figure} 
\includegraphics[width=0.45\textwidth,height=0.4\textwidth]{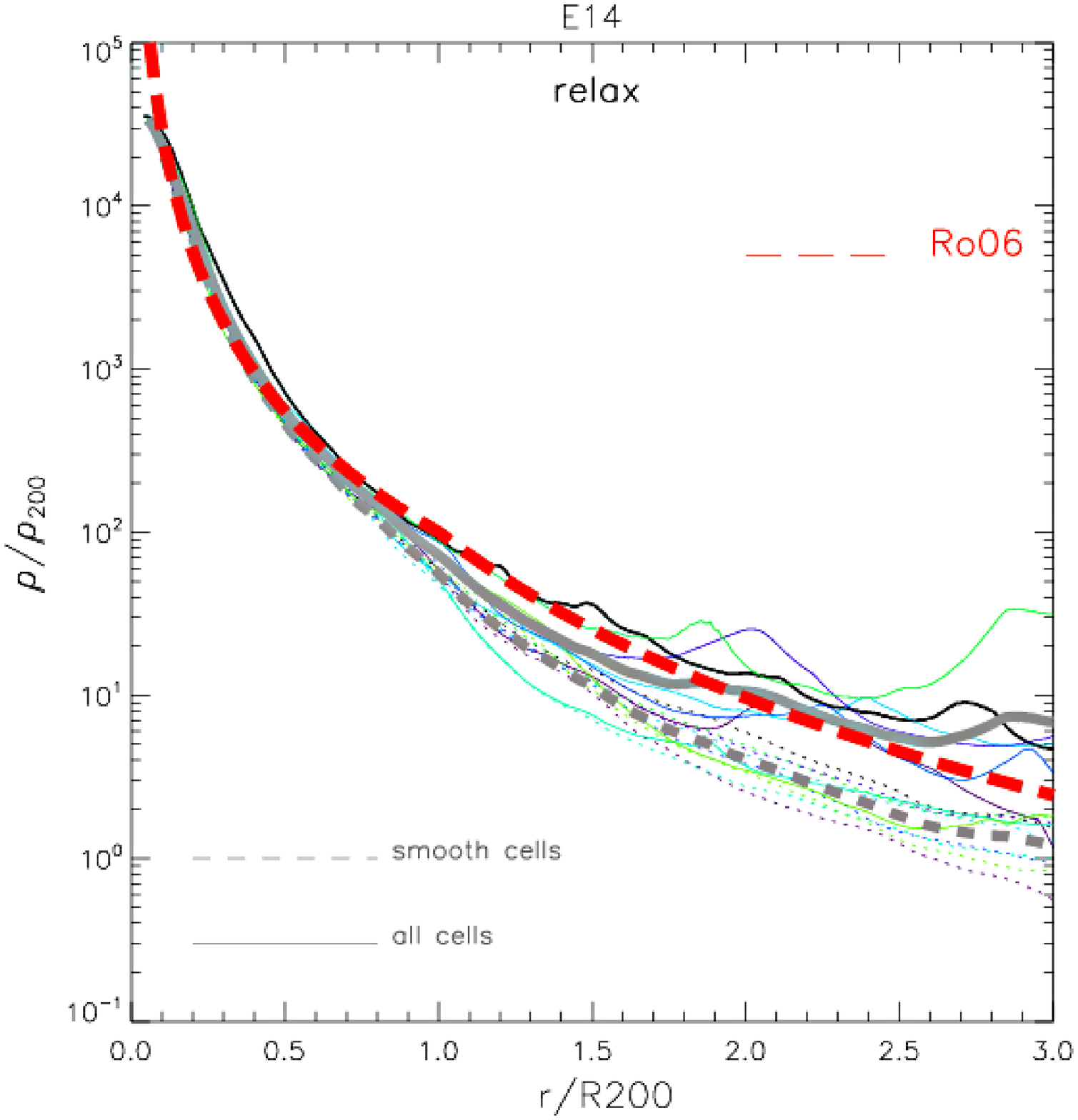}
\includegraphics[width=0.45\textwidth,height=0.4\textwidth]{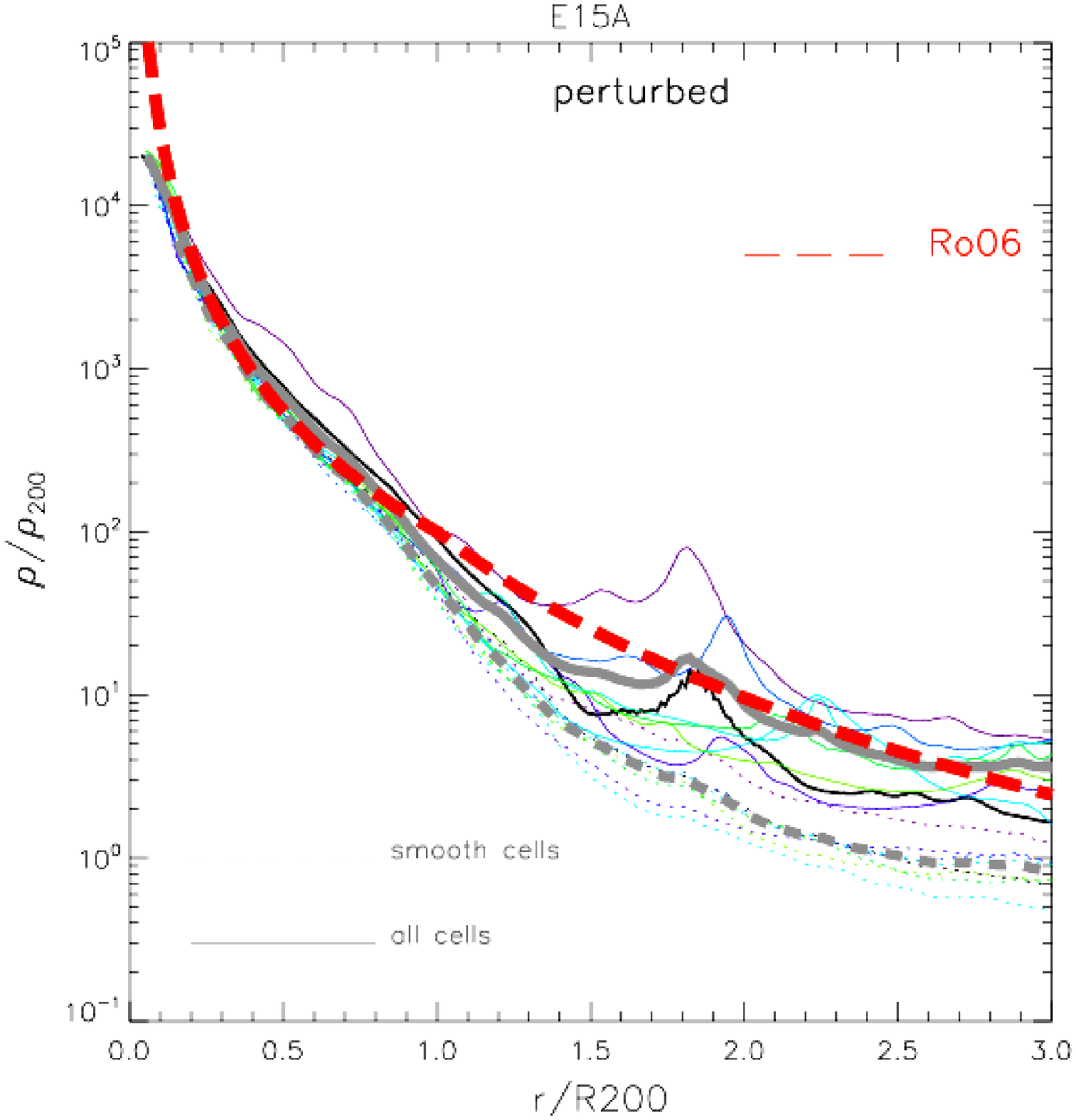}
\caption{3--D profiles of gas density for one perturbed ENZO cluster (top) and one relaxed ENZO cluster (bottom). 
Different colors
represent profiles for 8 different sectors, the solid lines are for the 
un-filtered cells while the dotted lines are for the $99$ per cent filtering. 
The red line in overlay shows the best-fit profile for the filtered data
of all GADGET2 clusters (from in Roncarelli et al.2006).}
\label{fig:prof_dens3}
\end{figure}

\begin{figure} 
\includegraphics[width=0.45\textwidth,height=0.37\textwidth]{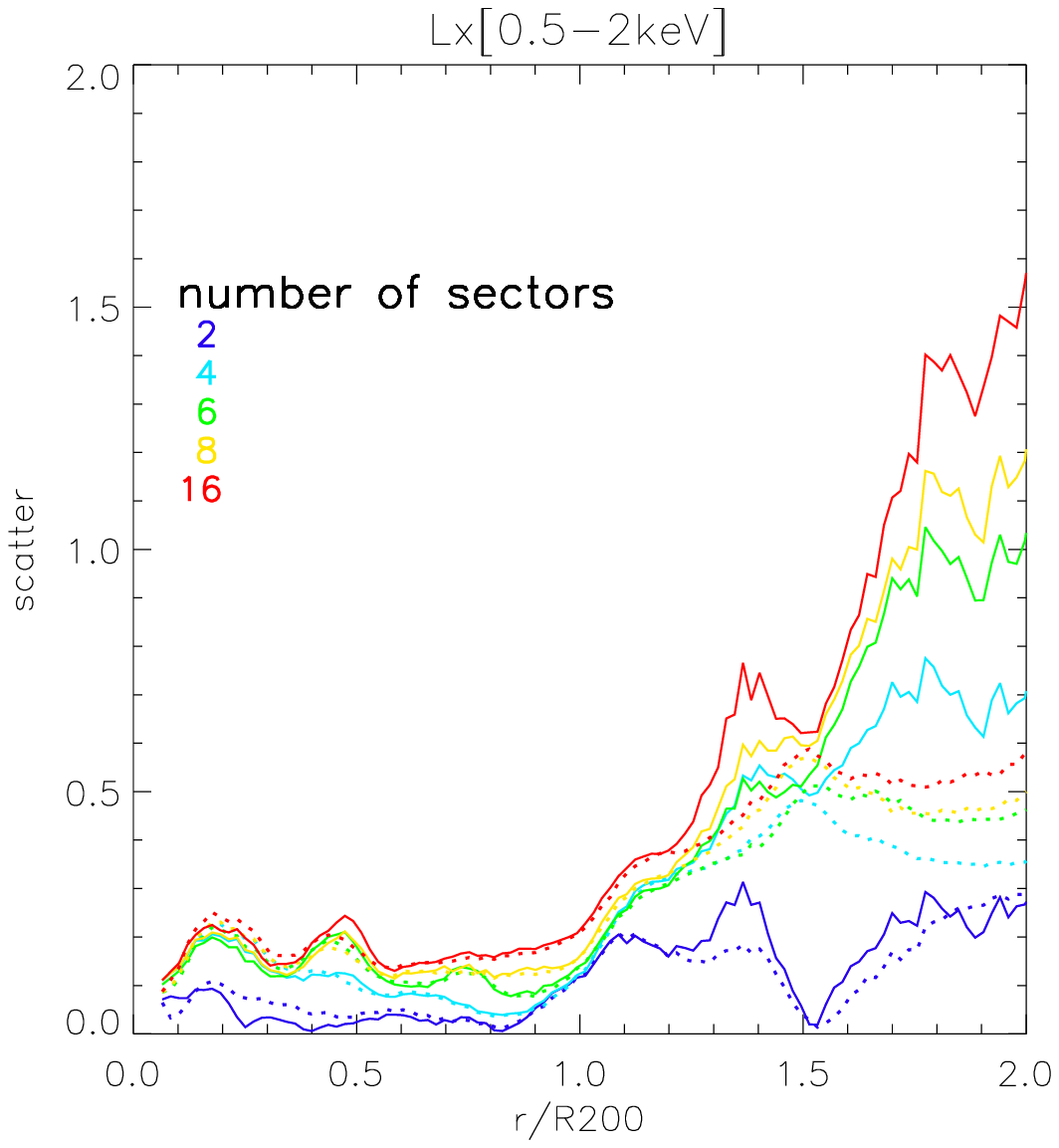}
\includegraphics[width=0.45\textwidth,height=0.37\textwidth]{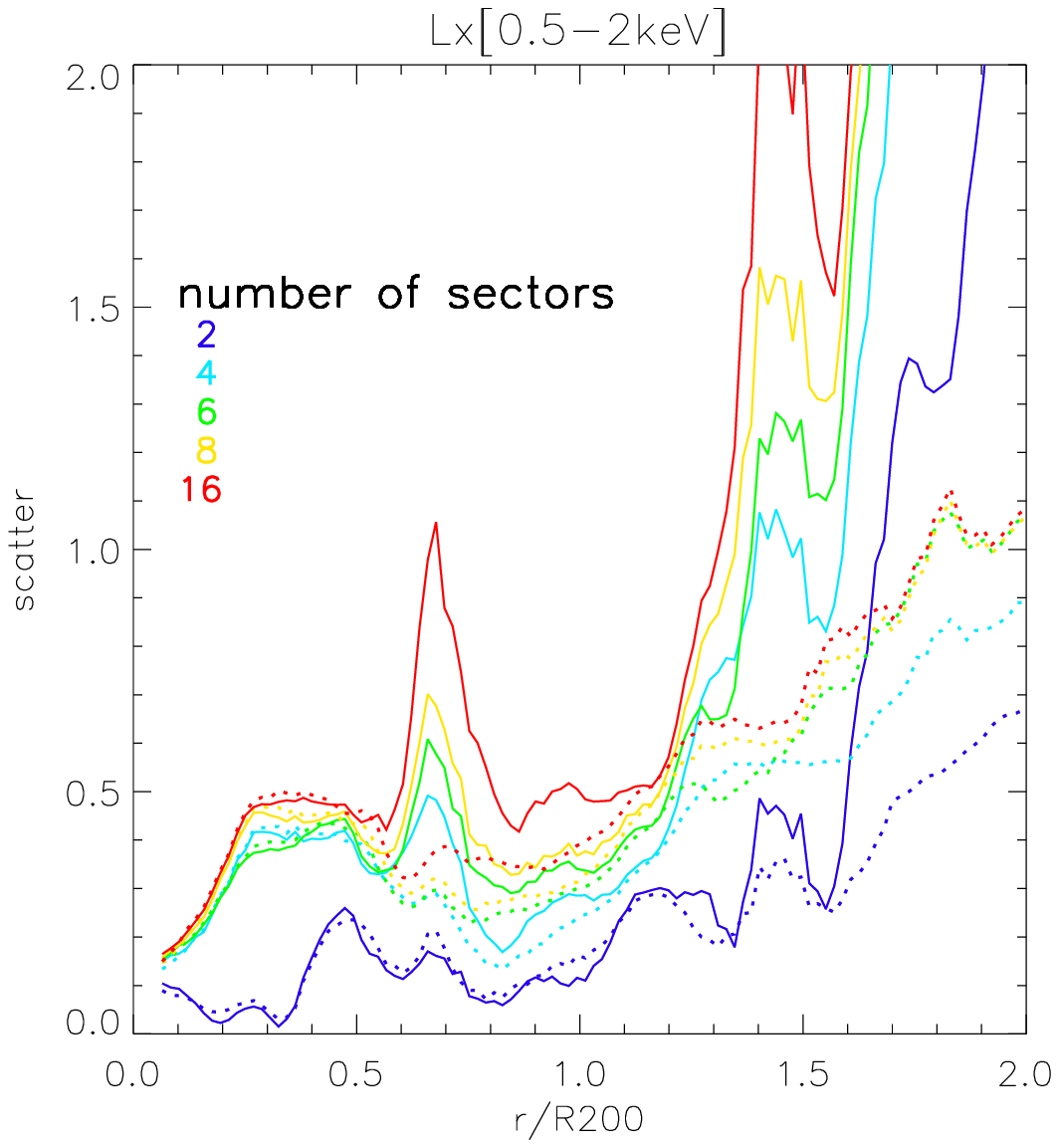}
\caption{Scatter profiles of projected $L_{X}$ at [0.5-2]keV for two representative clusters of sample (ENZO=top, GADGET2=bottom). 
The solid lines are for the un-filtered cells while the dotted lines are for the $99$ per cent filtering. Every color
refers to a different number $N$ of angular sectors in the projected volume.}
\label{fig:x_conv}
\end{figure}

\subsection{Data reduction.}
\label{subsec:method}

In the present work, we study the distribution of the ICM density,
temperature, entropy and surface brightness in our two sets of simulated
galaxy clusters by analyzing their radial profiles
in the range between $0.01 \leq r/R_{200} \leq 2 $.
 For both set of simulations, the radial profiles were computed within shells of equal spacing
 ($\approx 25kpc/h$) centered on the center of total (gas+DM) mass of each system.
The gas matter distribution in the cluster volume is usually characterized by the presence 
of very dense sub-clumps, with the typical dimension of galaxies or small galaxy-groups ($M 
< 10^{13} M_\odot$). 
In real X-ray observations, these dense and bright clumps can 
efficiently be masked out when present, and since in this work we aim at comparing with observational results, we applied a recipe to
 mimic the excision of such clumps from the simulations too.
 
We followed the procedure introduced in Ro06 and for every 3--D shell in
the cluster radial profile
we sorted the gas 
particles/gas density in decreasing volume order (thus in increasing density order). We next
 computed the profiles of gas density, of spectroscopic-like temperature ($T_{\rm SL}$, see Rasia et al.2005), of entropy and  of X-ray luminosity
in the [0.5-2]keV band considering only 
the particles/cells summing up to a fix percentage of the total
 volume of the shell. For consistency with Ro06,
this percentage was set to the $99$ per cent of the total volume of
each shell.  This essentially make sure that, at each radius, the first densest percentile of the 
gas distribution within each spherical shell is removed from the data-set, and replaced with the average value
at the same radius. Tests showed that the excision of this first percentile from the original data-set efficiently removes most of collapsed gas clumps at each radius. For further details we address the reader to the discussion in Ro06.
A similar filtering procedure was applied also directly to the 2--D maps of projected X-ray luminosities, by substituting the $1$ per cent most
luminous pixels in all maps with the average value at the corresponding radius.
Figure \ref{fig:lx_smooth} shows the application of this method for one projected X-ray map 
of one cluster from the ENZO sample; similar maps for the GADGET clusters can be found in Ro06.
In Figure \ref{fig:prof_temp} we report the gas density, $T_{\rm SL}$ and gas entropy (defined as 
$S=T_{\rm SL}/n_{e}^{2/3}$, where $n_{e}$ is the electron number density) for all clusters
in the two samples, with the $99$ per cent filtering (all profiles
are normalized at $R_{200}$ to highlight self-similarity).
Since the ENZO sample contains more clusters, larger statistical fluctuations are found in this data-set; nevertheless the average profiles obtained for the two numerical methods are in agreement within a $10-20$ percent for $r>0.2 \cdot R_{200}$. On the other hand, as expected sizable differences are found for the innermost cluster regions, $r<0.2 \cdot R_{200}$, in line with the findings reported and investigated by many different works in the literature (e.g. Frenk et al.1999; Voit et al.2005; Wadsley et al.2008; Mitchell et al.2009; Springel 2010; Vazza 2010). 
Discussing the above differences is far beyond the purpose of this work; 
we just notice that for our main goal here the most interesting regions under study are $>0.5 \cdot R_{200} $, where a general good consistency between the two codes is ensured. 
The effect of additional physical
processes not modeled in these runs (e.g. cooling, star formation, magnetic fields etc.)
is not expected to play an important role. For completeness, the issue of non-gravitational physics
in the simulations is investigated for a few representative clusters in the Appendix (see also Ro06 and discussion therein).

The simulated X-ray emission from our set of clusters 
was produced by adopting
a MEKAL emission model (e.g. Liedahl et al.1995) for 
each gas particle/cell, according to its
temperature and by assuming a {\it constant} gas 
metallicity of
$Z/Z_{\odot}=0.2$. This value is likely lower than
 the actual metallicity in the innermost clusters region (e.g. Leccardi \& Molendi 2008),
 however it can be considered a good  approximation 
to model the spectral energy distribution radiated by the ICM at the cluster outskirts,
once that the effect of metal cooling is considered.

In order to highlight the systematic effects due to the dynamical 
state of the clusters in our sample, we estimated the power ratios 
$P3/P0$, from the multipole decomposition of the X-ray surface 
brightness images,
versus the centroid shift, $w$, 
evaluated within $R_{500}$ (see Cassano et al.2010). 
We classified as "disturbed" systems those for which the values of 
$P3/P0>10^{-7}$ and $w>0.02$  were found in at least 2 of the 3 projected maps along the coordinate
axes, or as "relaxed" otherwise {\footnote{Since non-radiative runs tend to produce
too much concentrated cluster cores than real clusters, we expunged the $r<0.1R_{200}$ region
from our analysis.}}. 
 This classification method splits our two samples in the following way: in GADGET2 we obtain 5 relaxed systems and 4 perturbed systems, while in ENZO we obtain 11 relaxed systems and 9 perturbed systems. Compared
to the dynamical classifications introduced in Va10 for the ENZO sample (which were based on the matter accretion history inside the virial volume, and not on X-ray properties as in this work) we report that the "perturbed" class here almost perfectly overlap with the
class of "post-merger" systems there, while the "relaxed" class contains the "relaxing" and "merging" classes of Va10.

The key feature discussed in this paper is the {\it azimuthal} scatter in
the cluster profiles along different cluster sectors, as a function of the radial distance from 
the center of clusters. 
We mimicked an observational-like procedure and fixed a line of sight towards the cluster center, defining a cylinder of radius $2 \cdot R_{200}$ and height $4 \cdot R_{200}$
along the line of sight   {\footnote{In the Appendix we also show a test using the larger line of sight of $6 \cdot R_{200}$, showing
that our results on the azimuthal scatter of all physical quantities are remain unaffected. }} . The volume of each cylinder was divided by cutting $N$ ``sectors'' along the line of sight with angular extension $\Delta \alpha = 2\pi/N$.
This was made in order to approximately reproduce the usual decomposition of the FOV in {\it Suzaku} observations of the peripheral regions of galaxy clusters
(e.g.  $\Delta \alpha \approx \pi/2$ in George et al.2009).
As an example, we report in Fig.\ref{fig:prof_dens3} the behavior of the gas density profiles
for a relaxed and a perturbed ENZO cluster, along 8 different sectors of the projected volume,
with and without the "99 per cent filtering". For comparison we also overplot in 
within the same Figure the mean gas density profile derived in Ro06 after the "99 per cent filtering" method (dashed red line).
 In both cases, the profile in at least 2 sectors outside of $\sim R_{vir}$ shows quite large variations, reaching
a factor $\sim 10$ in the case of one particular sector in the perturbed cluster. These large fluctuations are
due to clumps and dense filaments crossing the virial radius, and when the "99 per cent filtering" is applied the agreement between the various sectors is generally better than a $\sim 20-30$ per cent at all radii. A qualitatively similar behavior is found also for the other physical quantities, and in GADGET2 runs.

The azimuthal scatter along the cluster radius, $S_c(r)$, is quantified as:
\begin{equation}
S_{c}(r)=\sqrt {\frac{1}{N} \cdot \sum_{i} {\frac{[y_{i}(r)-Y(r)]^{2}}{[Y(r)]^{2}}  }},
\label{eq:scatter}
\end{equation}
where $y_{i}(r)$ is the radial profile of a given quantity, taken inside a given
{\it i-}sector, and $Y(r)$ is the average profile taken from all the cluster
volume. In principle, functional fitting procedures and ensemble averages can be use to 
derive $Y(r)$ for each 
cluster, limiting the bias from substructures in the estimate of the  
average spherical profiles (e.g. Kawahara et al.2008). However, we want to make 
our treatment here the most straightforward as possible, in order to be readily useful to single-object 
observations, and 
therefore in this work $Y(r)$ is computed directly from the simulated cluster cells/particles (or pixels in the
case of $L_{X}$) after the "99 per cent filtering", and with the addition of a median smoothing filter 
along the radius.
Our tests confirm however that this issue is not very relevant for the problem discussed here, and
that a simple estimate of $Y(r)$ from the filtered data does not affect our statistical results on 
$S_{c}(r)$ in any practical way (see Sec.\ref{sec:results}).

The choice of the number $N$ of sectors is arbitrary, and may be tuned in 
order to mimic what done in current {\it Suzaku} observations.
 In the following, we will show examples of the average behavior of  $S_{c}(r)$ using
$N=2$, $=4$ and $=8$.
However, we find that a good convergence of $S_{c}(r)$ along the radius, for all analyzed quantities, is achieved adopting
$N \geq 8$ for both the dynamical classes considered. 
 As an example, we show in Fig.\ref{fig:x_conv} 
the azimuthal scatter profiles for the projected X-ray projected luminosity at [0.5-2]keV for a cluster of each
sample. In the "filtered" case, at all radii $S_{c}(r)$ do not change by more than a $\sim 5$ per cent if the number
of sectors is increased from $N=8$ to $N=16$. In the un-filtered case, however, the presence of the point-like bright contribution by collapsed gas clumps leave strong fluctuations in the azimuthal scatter even at the small
angular scale of $\Delta \alpha=\pi/8$ sampled taking $N=16$.   
This may serve as a guidance for future X-ray observations of the outermost 
regions of galaxy clusters: {\it an accurate reconstruction
of the thermal gas distribution of galaxy clusters is
reached with good accuracy for sectors of angular size 
$\Delta \alpha \sim \pi/4$ (or smaller), provided that an efficient masking of point-like contribution from
dense gas clumps is performed.}

\begin{figure*} 
\includegraphics[width=0.45\textwidth,height=0.35\textwidth]{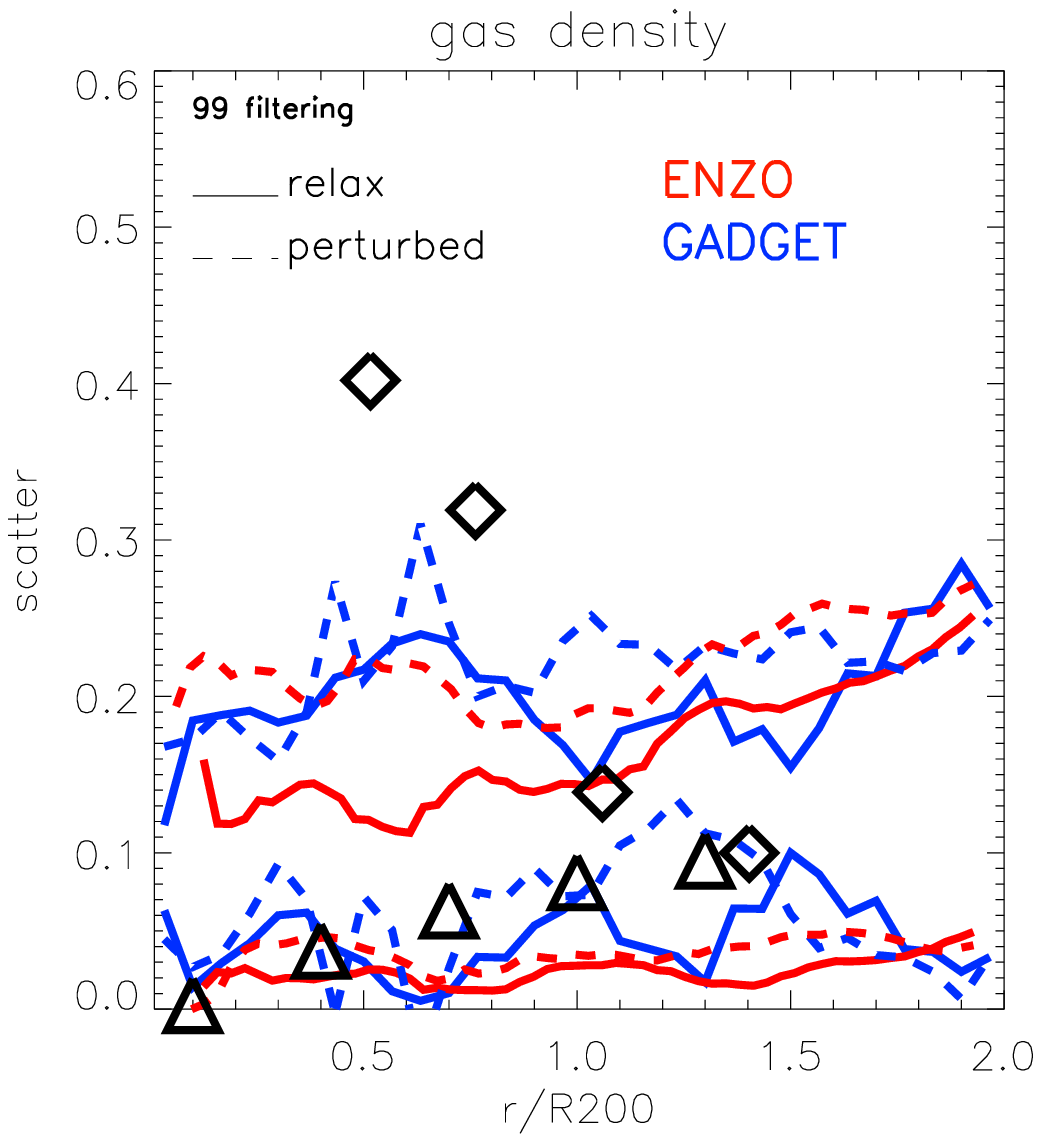}
\includegraphics[width=0.45\textwidth,height=0.35\textwidth]{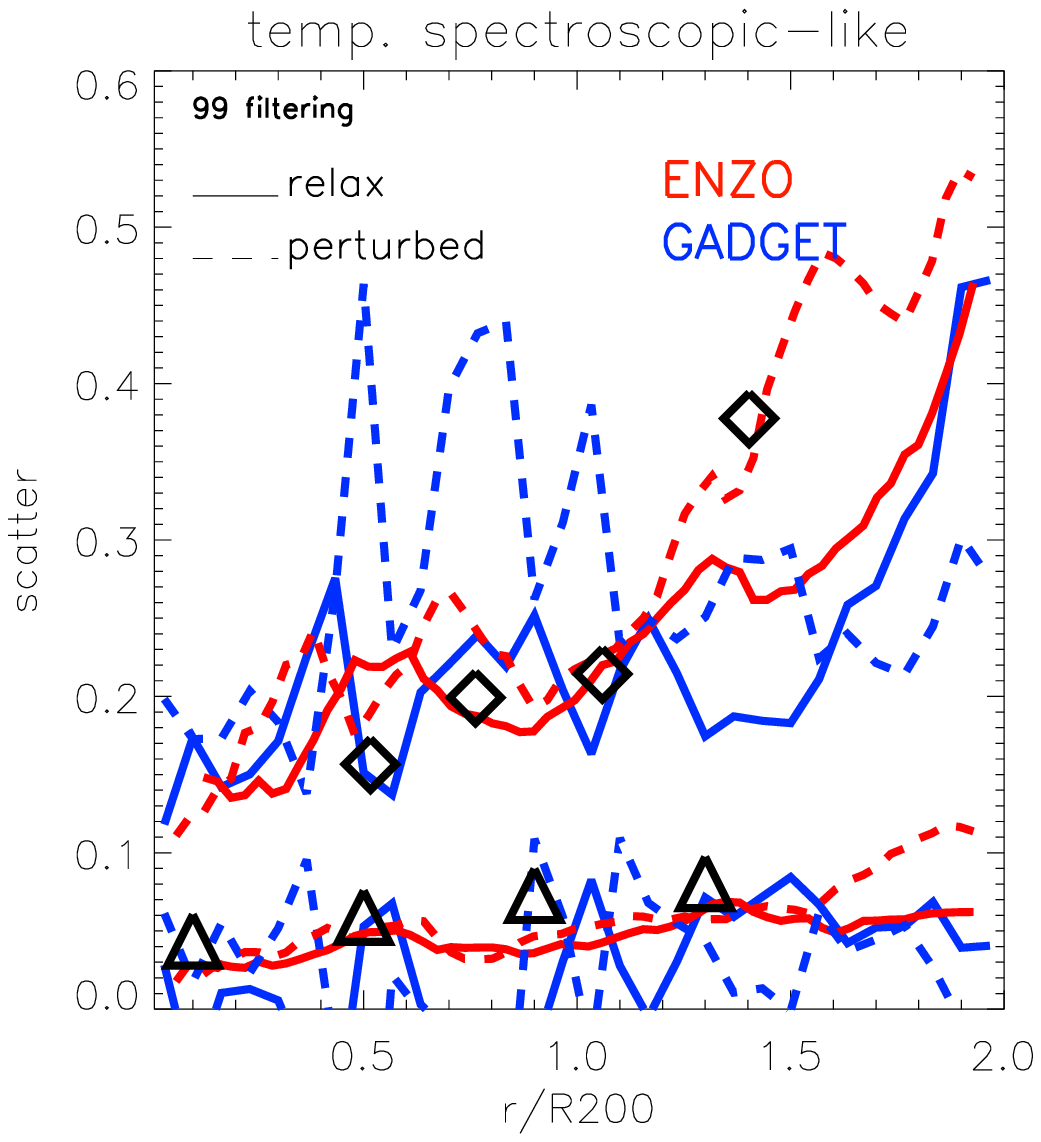}
\includegraphics[width=0.45\textwidth,height=0.35\textwidth]{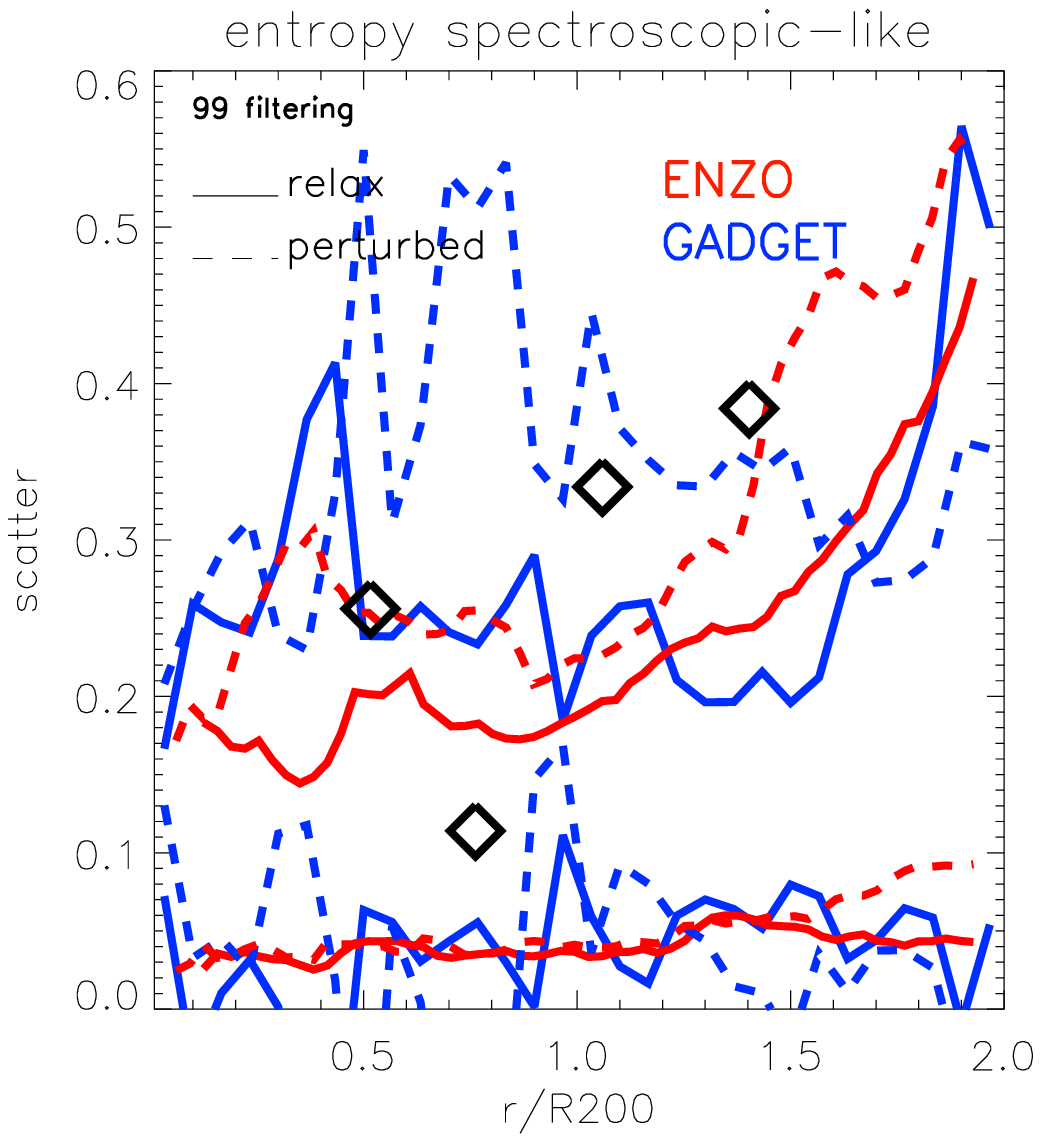}
\includegraphics[width=0.45\textwidth,height=0.35\textwidth]{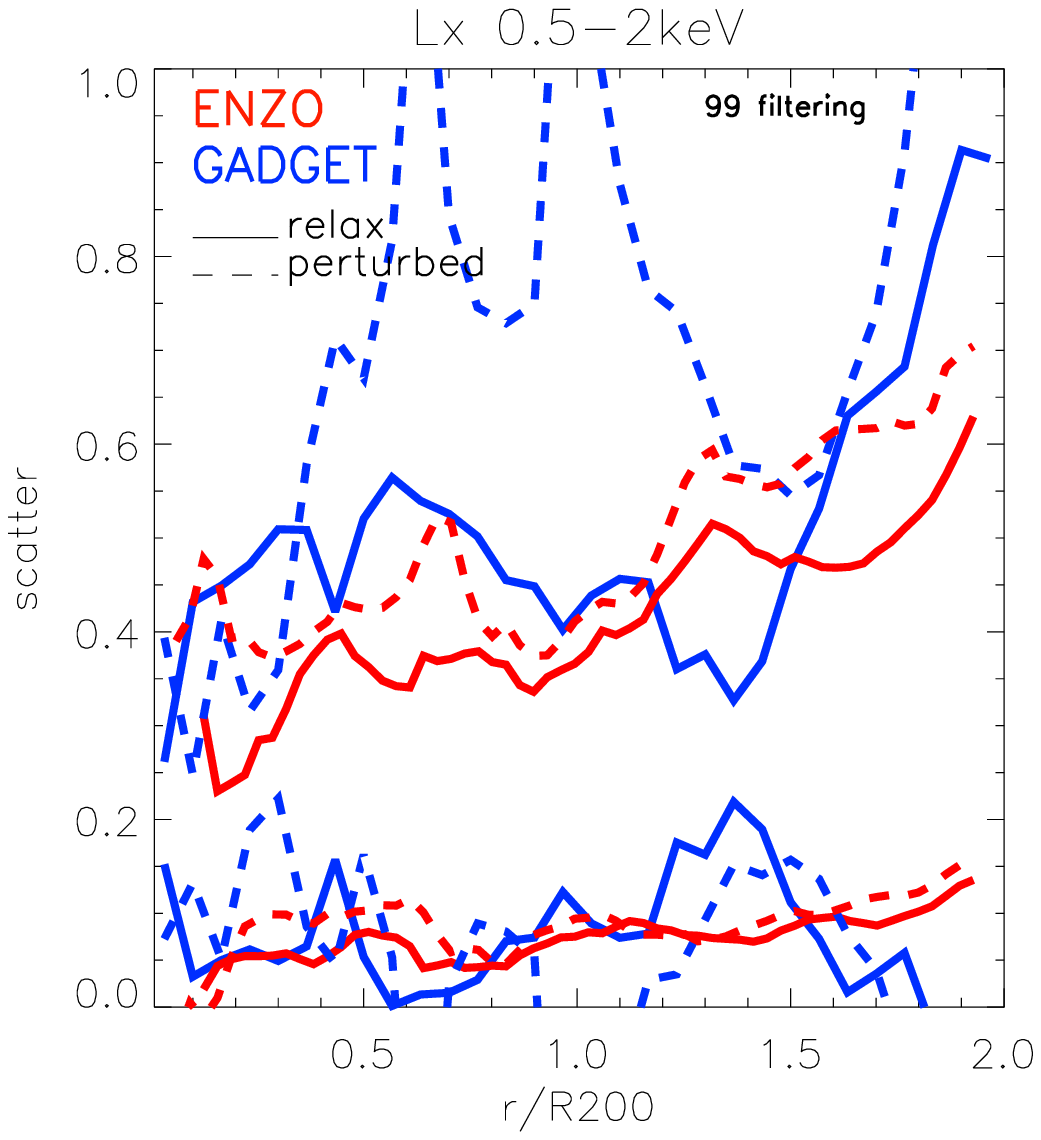}
\caption{Average scatter profiles for all clusters of the paper (ENZO=red; GADGET=blue) for the data with ``99 percent filtering'' and for $N=8$ sectors. 
For each cluster sample we show the 2 lines enclosing the $\pm 1 \sigma$ around the mean azimuthal scatter.
 The solid lines are for the ``relaxed'' clusters while the dashed lines are for the ``perturbed'' clusters.
 Overlayed are the observational {\it Suzaku} point of George et al.(2009) for PKS0745-191 (diamonds), and of Bautz et al.(2009) for A1795 (triangles).}
\label{fig:prof_scatter}
\end{figure*}
\section{Results}
\label{sec:results}

 Our total sample of 29 clusters, together with the filtering procedure outlined in the previous Section, allows us 
to assess the  intrinsic level of azimuthal scatter expected for the clusters profiles, as an effect of
the hierarchical clustering and cosmic evolution process driven by gravitational processes alone.

In Figure \ref{fig:prof_scatter} we report our results for the 
azimuthal scatter profiles of the galaxy clusters simulated in the 2 codes, by adopting
the preferred value of $N=8$ sectors and the ``99 per cent filtering'' (see Sec.\ref{subsec:method}). 
 Each sample was also divided into perturbed and relaxed systems (see different line-styles in Fig.\ref{fig:prof_scatter}.
Each couple of lines with the same style shows the contours of $S_{c}(r) \pm \sigma_{S_{c}(r)}$ (the standard deviation is computed across for each dynamical sub-sample of the cluster sample). 
The filtered profiles show an agreement between the two codes better than a $\sim 5$ per cent at any radius when
relaxed systems are considered, and slightly larger differences for the case of perturbed systems.
In general, we report that for gas density, gas temperature and gas entropy the intrinsic mean azimuthal
scatter computed for $N=8$ sectors is $S_{c}(r) \sim 10 \pm 5$ per cent for $r<R_{200}$ for both codes,
and  $S_{c}(r) \sim 25 \pm 10$ for the X-ray luminosity at [0.5-2]keV within the same radius. Outside of $R_{200}$, both 
codes present a gradual increase of the azimuthal scatter in all quantities, up to a factor $\sim 2$ at
$2 \cdot R_{200}$. In general, pertubed objects present larger azimuthal scatter at all radii, $\sim 20-40$ percent more compared to the scatter measured in relaxed systems. This is explained noticing that, even if clumps are efficiently removed by our filtering procedure, perturbed structures are generally characterized by a larger amount of chaotic motions and shocks at all radii (e.g. Va10 for an analysis of shocks in the ENZO sample).
An important point to notice here is that the agreement between the azimuthal scatter in two codes is  generally better than the agreement between the simple radial profiles (Fig.\ref{fig:prof_temp}).   This suggests that, despite of the small but non-negligible differences in the 
cluster profiles of the GADGET2 and ENZO samples, which can be due to a number of possible reasons (i.e. slightly difference
cosmological setup, cluster extractions procedures, numerical issues etc.), our predictions on the level
of intrinsic azimuthal scatter in clusters are robust and
independent on the particular code adopted.

It is now interesting to compare these findings with the (few) available observations of azimuthal scatter derived
from long  {\it Suzaku} exposures of PKS0745-191 (George et al.2009, black diamonds in Fig.\ref{fig:prof_scatter}) and of A1795 (Bautz et al.2009, black triangles).
The first observation was performed with $\Delta \alpha \approx \pi/2$ (thus $N=4$), while
for the second  $\Delta \alpha \approx \pi$ ($N=2)$.

 In general, the reported azimuthal scatter of PKS0745-191 lies at the uppermost part of our average distributions in Fig.\ref{fig:prof_scatter}, while the azimuthal scatter of A1795 lies in the lowermost part. 
Even if the radial sampling available to observations is yet quite scarce, it seems that the trends with radius
of the available data can be reproduced by cosmological simulations, with the exception of the peculiar {\it decreasing}  trend with radius reported for the gas density of PKS0745-191. 

\begin{figure}
\includegraphics[width=0.235\textwidth]{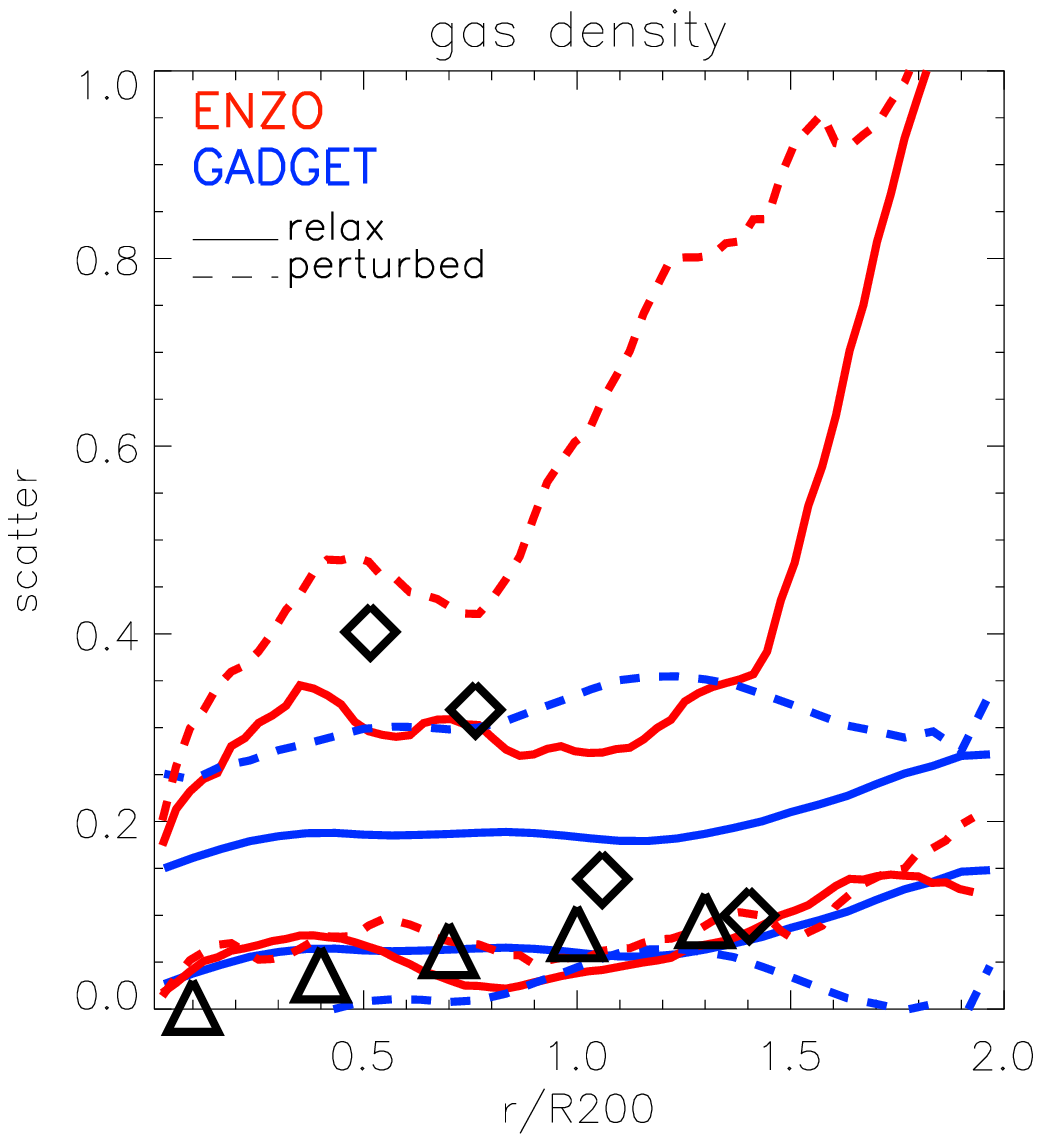}
\includegraphics[width=0.235\textwidth]{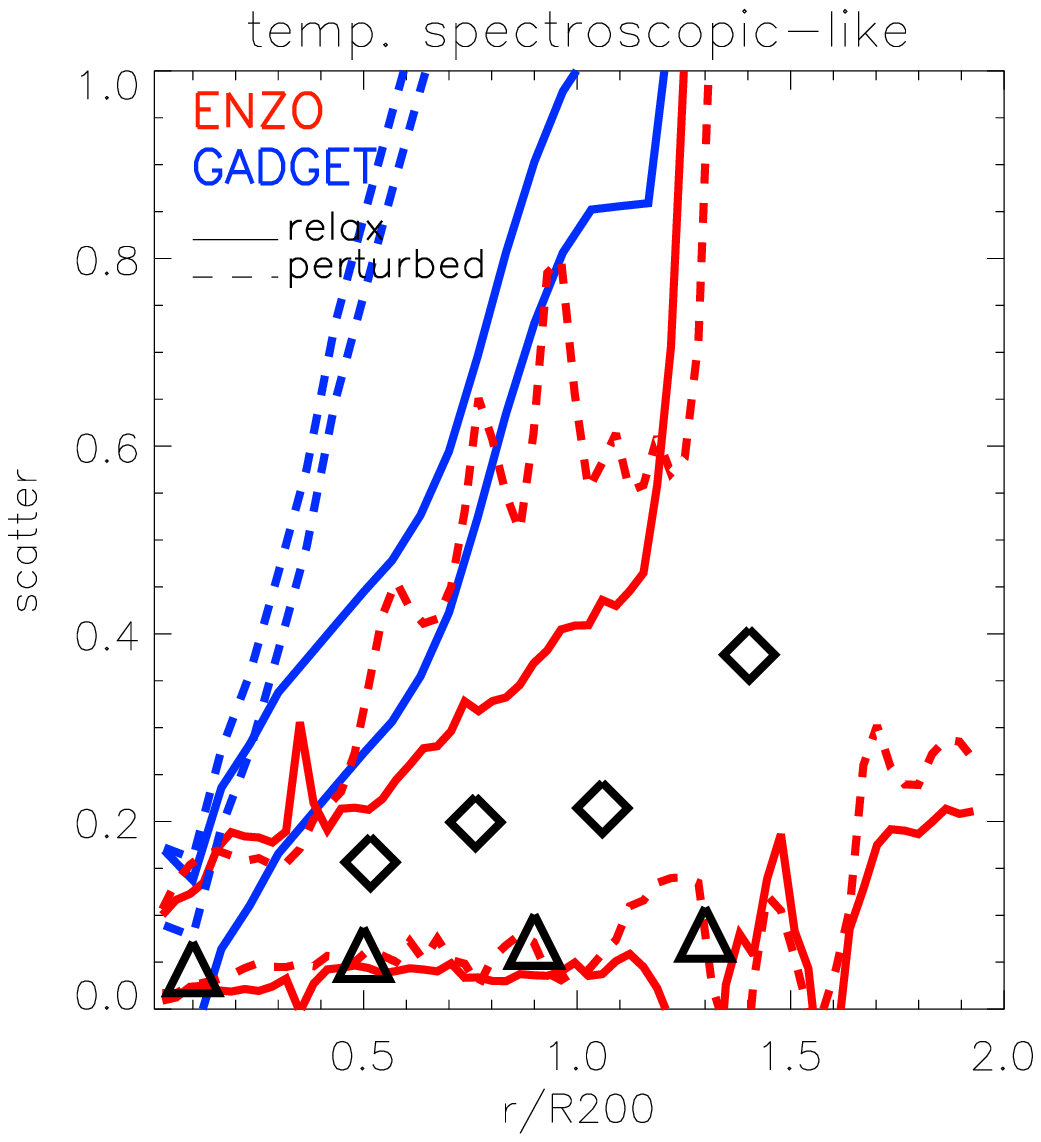}
\caption{Same as in Figure \ref{fig:prof_scatter}, but for the un-filtered gas density and gas temperarure and for $N=8$.}
\label{fig:prof_scatter2}
\end{figure}

At least two reasons may be responsible for the larger azimuthal scatter of PKS0745-191 for $r<R_{200}$ compared to our simulated results. First, PKS0745-191 is a system affected by a un-usually large perturbation (caused for instance by a merger or by the accretion of a large scale filament), which is observed only in 
a few exceptional cases in our sample of simulated clusters. 
We notice, however, that PKS0745-191 is usually classified as a relaxed cool-core cluster 
(George et al.2009 and references therein); anyway this classification is just based on the 
cluster morphology around the core region, and thus might be not at variance with our definition of "perturbed" systems, for which the whole volume of the cluster is analyzed.
A second possibility, is that some un-resolved substructures are present in one of the 4 FOV targeted by {\it Suzaku}, a possiblity that the authors themselves considered in their work (George et al.2009). Indeed, if the azimuthal scatter is computed for the un-filtered gas 
density and $T_{\rm SL}$ (see Fig.\ref{fig:prof_scatter2}) for all the simulations in our data-set, much larger scatters, compatible with the findings of George et al.(2009), are found. In the case  
 of ENZO 
runs, moreover, it seems that on average also the ``inverted'' trend in gas density for $0.5 \leq r/R_{200} \leq 1 $ is statistically reproduced by the simulated data, due to the effect of resolved filaments of gas matter impinging
 at the cluster periphery along some preferential sectors, which are well resolved 
with the mesh refinement strategy adopted in this particular data-set (see Va10 for details). Similar results were also reported by Burns et al.(2010).
  
\begin{figure}
\includegraphics[width=0.235\textwidth]{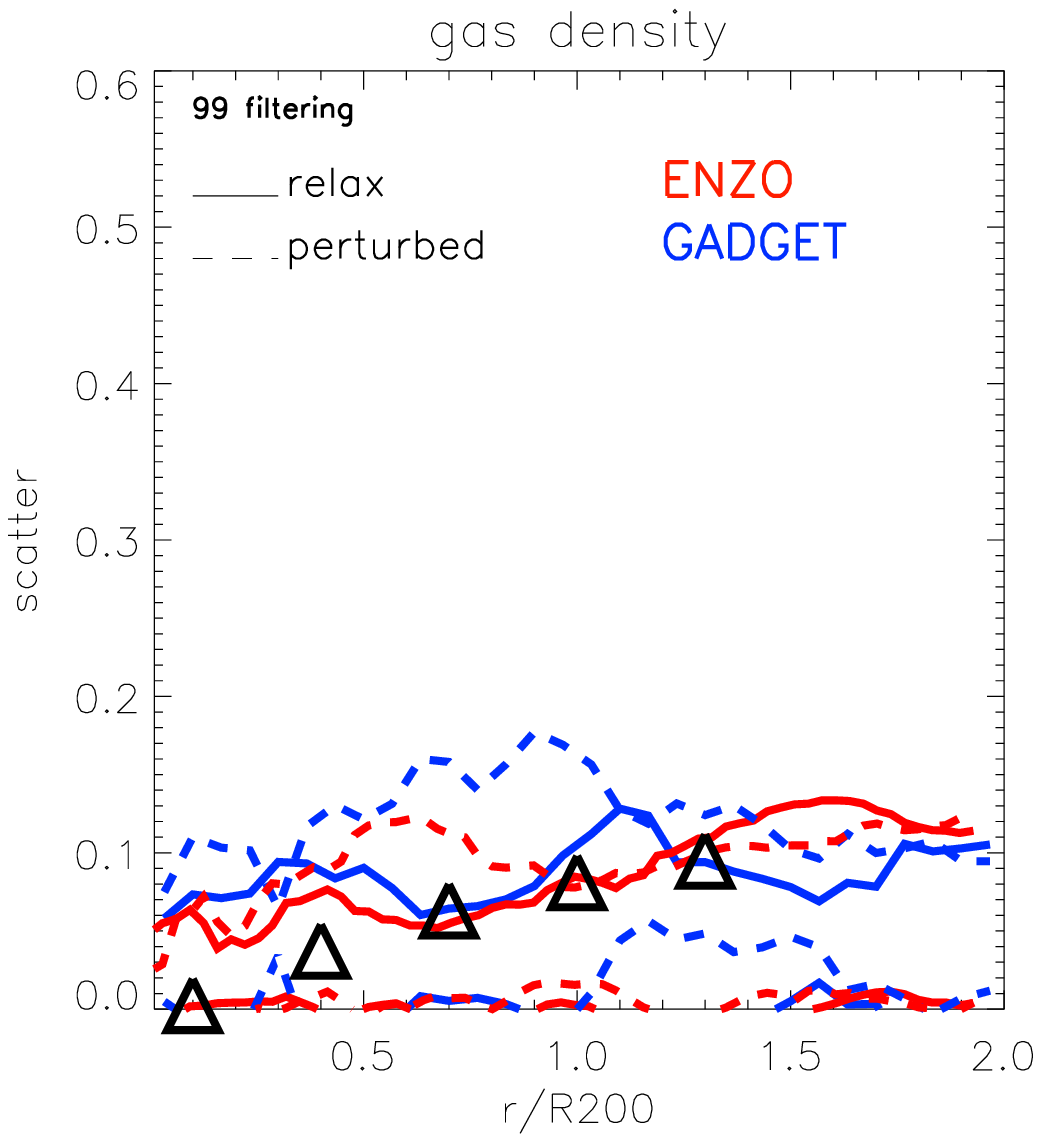}
\includegraphics[width=0.235\textwidth]{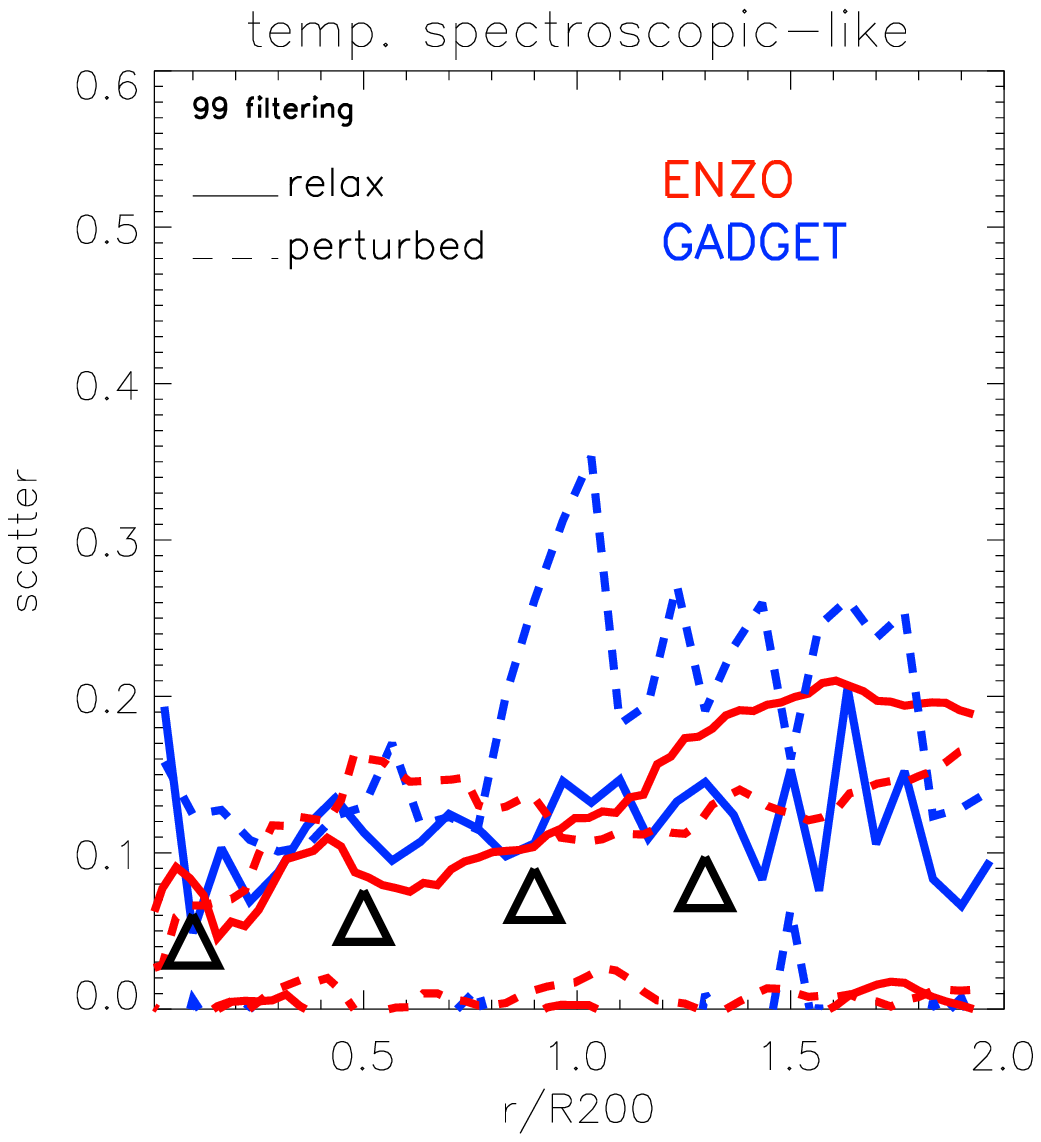}
\includegraphics[width=0.235\textwidth]{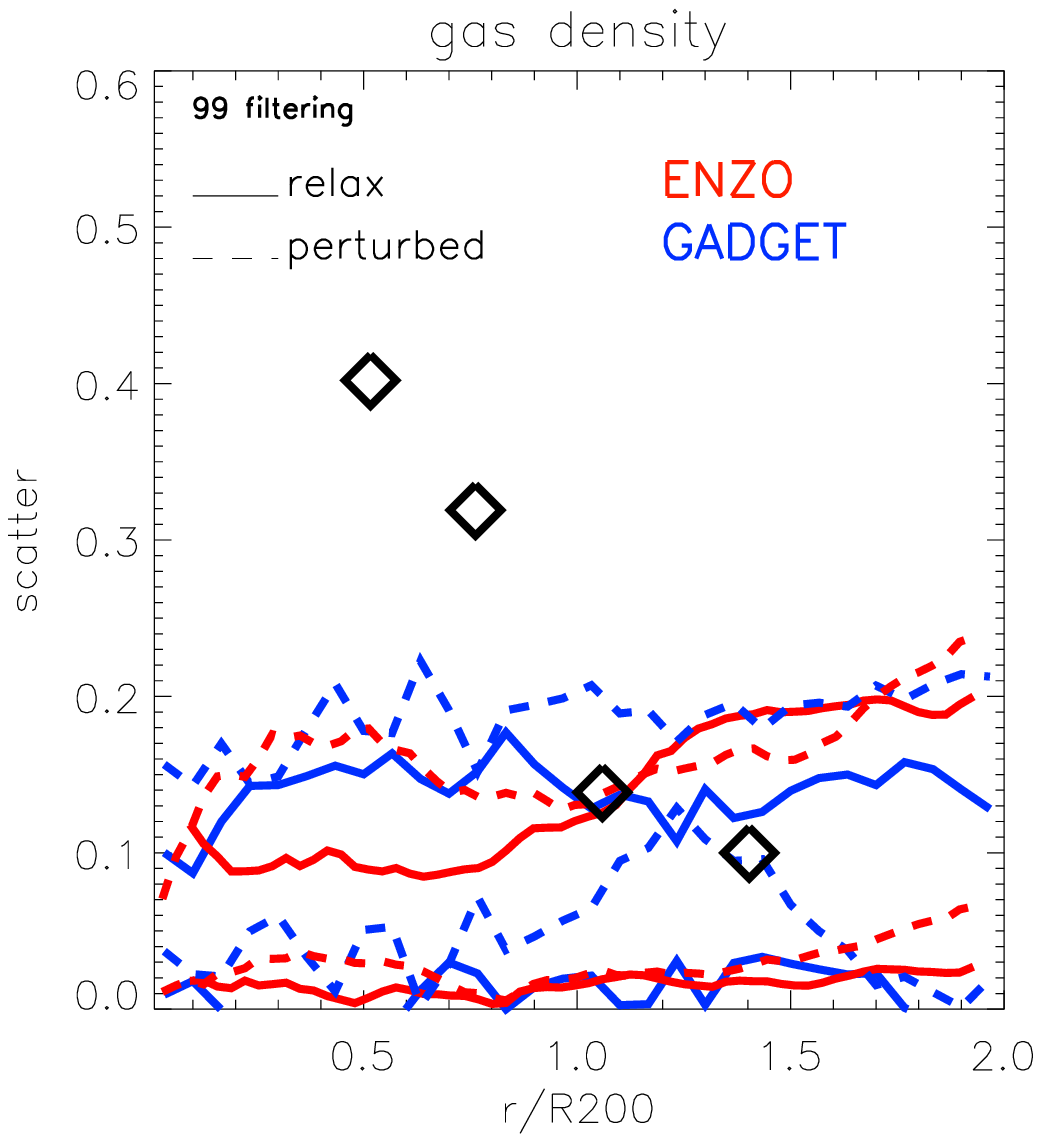}
\includegraphics[width=0.235\textwidth]{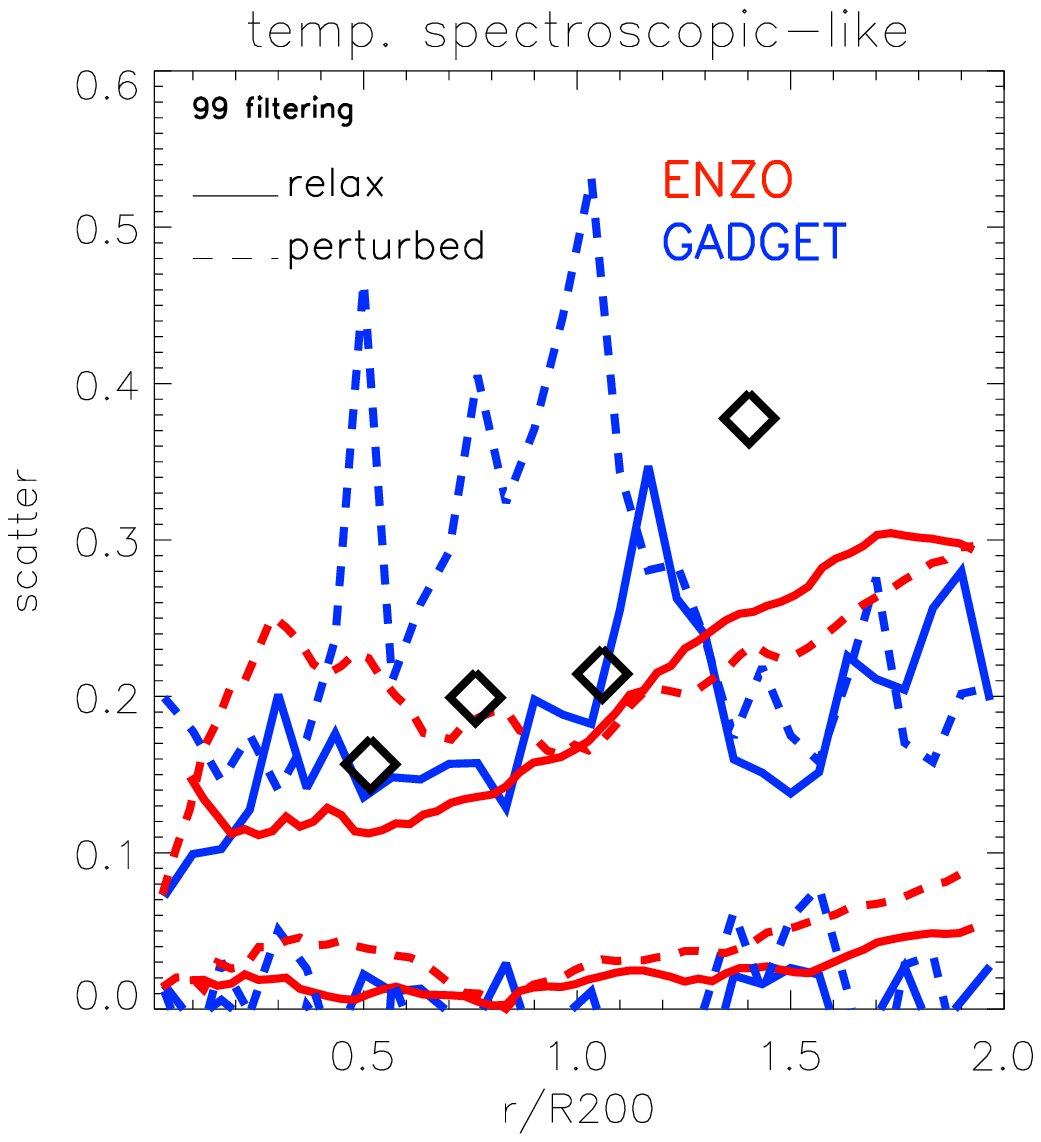}
\caption{Azimuthal scatter profiles for gas density and gas temperature for the same samples of Fig.\ref{fig:prof_scatter}, for N=2 (top panels) and N=4 (bottom panels). The meaning of colors and linestyles is an in Fig.\ref{fig:prof_scatter}. The points derived from A1795 (Bautz et al.2009) are reported in the top panels (as triangles), while the points derived from PKS0745-191 (George et al.2009) are reported in the bottom panels (as squares).}
\label{fig:prof_scatter3}
\end{figure}

This test compared the "intrinsic" azimuthal scatter measured in our filtered simulations (using a number of sectors which produces a convergence in the estimate of $S_{c}(r)$) with the presently best available data from real {\it Suzaku} observations; however the data from George et al.(2009) were derived from $N=4$ sectors, and the
data from Bautz et al.(2009) were derived from $N=2$. As we shown in Fig.\ref{fig:x_conv}, the use of different angular sampling is expected to affect the estimate of the real $S_{c}(r)$ for $\Delta \alpha > \pi/4$.  
To investigate this effect, we also repeated the measures of the azimuthal scatter profiles across the whole
sample of simulations, assuming $N=2$ (top panels of Fig.\ref{fig:prof_scatter3}) and $N=4$ (bottom panels of Fig.\ref{fig:prof_scatter3}). 
As expected from the tests shown in Sec.\ref{subsec:method}, the azimuthal scatters
are progressively reduced at all radii as the angular size of sectors is increased. With $N=2$ our measurement of the average azimuthal scatter in clusters is  now well in agrement with the data obtained for A1795 from Bautz et al.(2009) observations. In the case of $N=4$ still the scatter in gas density for PKS0745-191 from George et al.(2009), derived with the same number of sectors, can be explained by our data only assuming the contribution from un-resolved gas clumps.

\section{Discussion and conclusions}
\label{sec:conclusions}

In this work we report upon the first statistical measure of the azimuthal scatter in the thermal distribution of galaxy clusters (analyzing gas density,
gas temperature, gas entropy and X-ray emission) with a large set of cosmological simulations run with ENZO (Vazza 
et al.2010) and with GADGET2 (Roncarelli et al.2006).
 This joint sample of simulations offers a unique view to statistically constrain the intrinsic level of azimuthal scatter 
in the cluster observables, as produced by the simple gravitational processes
which follow the hierarchical growth of
cosmic structures.This can be compared to the (few) real data recently obtained with long {\it Suzaku} exposures
of specific sectors in nearby clusters (e.g. PKS0745-191 by George et al.2009; A1795 by Bautz et al.2009).
For this purpose, we apply a post-processing technique to our simulations designed to remove the dense collapsed
gas clumps which are usually masked out from real X-ray observations (Roncarelli et al.2006).
According to our study, a statistically  convergent
 modeling of the intrinsic azimuthal scatter in cluster profiles is achieved for sectors of angular
size $\Delta \alpha \approx \pi/4$ or smaller, almost regardless of the numerical scheme and of the dynamical state of clusters.
 The level of azimuthal scatter reported for the available observations (George et al.2009; Bautz et al.2009) 
can be in principle well reconciled with our simulations when using a similar number of angular sectors (with angular
size $\Delta \alpha =\pi/2$ for PKS0745-191 and  $\Delta \alpha =\pi$ for A1795); however the peculiar
trend of gas density for $r<R_{200}$ in PKS0745-191 may be well reconciled with simulations only in the case that some 
un-resolved gas substructure (e.g. a filament or a gas clump) is be present in one of the {\it Suzaku} FOVs.

To summarize our results (for simplicity restricting to gas density and spectroscopic-like temperature), we report in Table 2 our estimates of the azimuthal scatters in our sample of clusters, after the ``99 percent filtering'' and
assuming $N=8$ and $N=4$.

In the future, a larger sample of real cluster observations would be desirable in order
 to fully account for the variety of dynamical states present in clusters, which may affect the observed azimuthal scatter in a sizable ways.

We also comment that, at the present stage, the effect of non-gravitational
 physical processes acting at the cluster peripheral regions (e.g. AGN-feedback, magnetic
 fields, Cosmic Rays, etc.) is not required to reproduce the presently available observations, given that present cosmological
 high-resolution and "simple" non-radiative runs are suitable to reproduce the trends reported by observations (see also the 
tests in the Appendix). This findings are qualitative in line with previous results discussed by Romeo et al.(2006); Roncarelli et al.(2006); Burns et al.(2010).

However, in the future a self-consistent modeling
of the ICM metallicity and chemical enrichment processes in the cluster outskirts may be important to model the X-ray
luminosity in a more realistic way (e.g. by adding chemical evolution, transport of metals and line cooling).

\begin{table}
\label{tab:tab2}
\caption{Reference values for the azimuthal scatter of gas density and spectroscopic like temperature for the GADGET2 and ENZO samples, using the "99 percent filtering" at 3 different radius from the center, assuming $N=4$ and $N=8$. 
The mean azimuthal scatter values are reported with their $\pm\sigma$ standard error. The last column shows the dynamical state of the cluster sub-samples considered ({\it R=}relaxed, {\it P=}perturbed). }
\begin{tabular}{c|c|c|c|c}
$\rho_{gas}$, {\it  N=8} & $0.5 R_{200}$ & $R_{200}$ & $1.5 R_{200}$ & {\it state}\\ 
\hline
{\rm GADGET} & $0.124   \pm 0.093$ & $0.116 \pm  0.053$ & $ 0.133  \pm 0.047$ &R\\
                          & $0.141 \pm  0.069$ & $0.154 \pm  0.082$ & $0.141  \pm 0.102$ &P\\

{\rm ENZO} &  $0.075 \pm   0.054$  & $ 0.081 \pm  0.054$ & $ 0.103 \pm  0.086$ &R\\
                &  $0.116 \pm   0.071 $ & $ 0.104 \pm 0.072$ &  $0.139 \pm  0.096$ &P \\
\hline
$\rho_{gas}$, {\it  N=4} & $0.5 R_{200}$ & $R_{200}$ & $1.5 R_{200}$ & {\it state}\\
\hline
{\rm GADGET} & $ 0.074 \pm  0.077$ & $0.081 \pm  0.061$ & $0.087  \pm 0.061$ & R \\
                        & $0.114 \pm 0.064$ & $0.128 \pm 0.071$ &  $ 0.123 \pm 0.073$ & P \\
{\rm ENZO} & $0.055 \pm 0.047$ & $0.065 \pm  0.051$ & $0.104 \pm  0.087$ & R \\ 
                   & $ 0.100 \pm 0.068$ & $0.076 \pm 0.057$ & $ 0.097 \pm  0.065$ & P \\
\hline
$T_{\rm SL}$, {\it N=8} & $0.5 R_{200}$ & $R_{200}$ & $1.5 R_{200}$ & {\it state}\\
\hline
{\rm GADGET} &   $0.102 \pm   0.049$ & $ 0.117 \pm  0.087$ & $0.139 \pm  0.072$ & R\\
                    & $0.122 \pm 0.341$ & $ 0.185 \pm  0.127$ & $ 0.143 \pm   0.081$ & P \\
{\rm ENZO}& $0.109 \pm  0.073$ & $ 0.111 \pm 0.073$ & $ 0.152 \pm  0.095$ & R \\
                   & $ 0.127 \pm   0.076$  & $0.119 \pm  0.073$   &  $ 0.216 \pm  0.148$ & P \\
\hline
$T_{\rm SL}$,  {\it N=4} & $0.5 R_{200}$ & $R_{200}$ & $1.5 R_{200}$ &  {\it state}\\
\hline
{\rm GADGET} &$ 0.079 \pm 0.057$  & $0.091  \pm  0.097$  &  $0.087 \pm  0.065$ & R \\ 
                      & $0.110 \pm 0.356$ &  $ 0.166 \pm 0.277$ & $ 0.118 \pm 0.040$ & P \\
{\rm ENZO} &  $0.068 \pm 0.061$  & $0.087 \pm 0.072$  &   $0.143 \pm 0.116$ & R \\
                   & $0.128 \pm   0.086$   & $0.092 \pm    0.074$    & $0.138 \pm  0.089$ & P \\
 \end{tabular}
\end{table}

 \section*{Acknowledgements}
We thank C.Gheller and R.Brunino for their priceless support at HPC-CINECA (Italy), and G.Brunetti of fruitful discussions.
We acknowledge the help of D.Fabjan in the treatment of GADGET simulations data.
F.V. acknowledges the 
usage of computational resources under the CINECA-INAF 2008-2010 agreement
and the 2009 Key Project ``Turbulence, shocks and cosmic rays electrons 
in massive galaxy clusters at high resolution'', and partial financial support from the grant FOR1254 from Deutschen Forschungsgemeinschaft.  S.E. and M.R. acknowledge the financial contribution from contracts ASI-INAF I/023/05/0 and I/088/06/0. M. R. acknowledges the support of grant ANR-06-JCJC-0141. K.D. acknowledges the support by the DFG Priority Programme 1177 and additional support by the DFG Cluster of Excellence "Origin and Structure of the Universe".

\bigskip
\bigskip

\appendix

 \section*{Appendix}

We briefly present  here additional tests to assess the reliability of the results against a few numerical effects that might play a role in the computation of the average azimuthal scatter.

\begin{figure}
\begin{center}
\includegraphics[width=0.235\textwidth,height=0.23\textwidth]{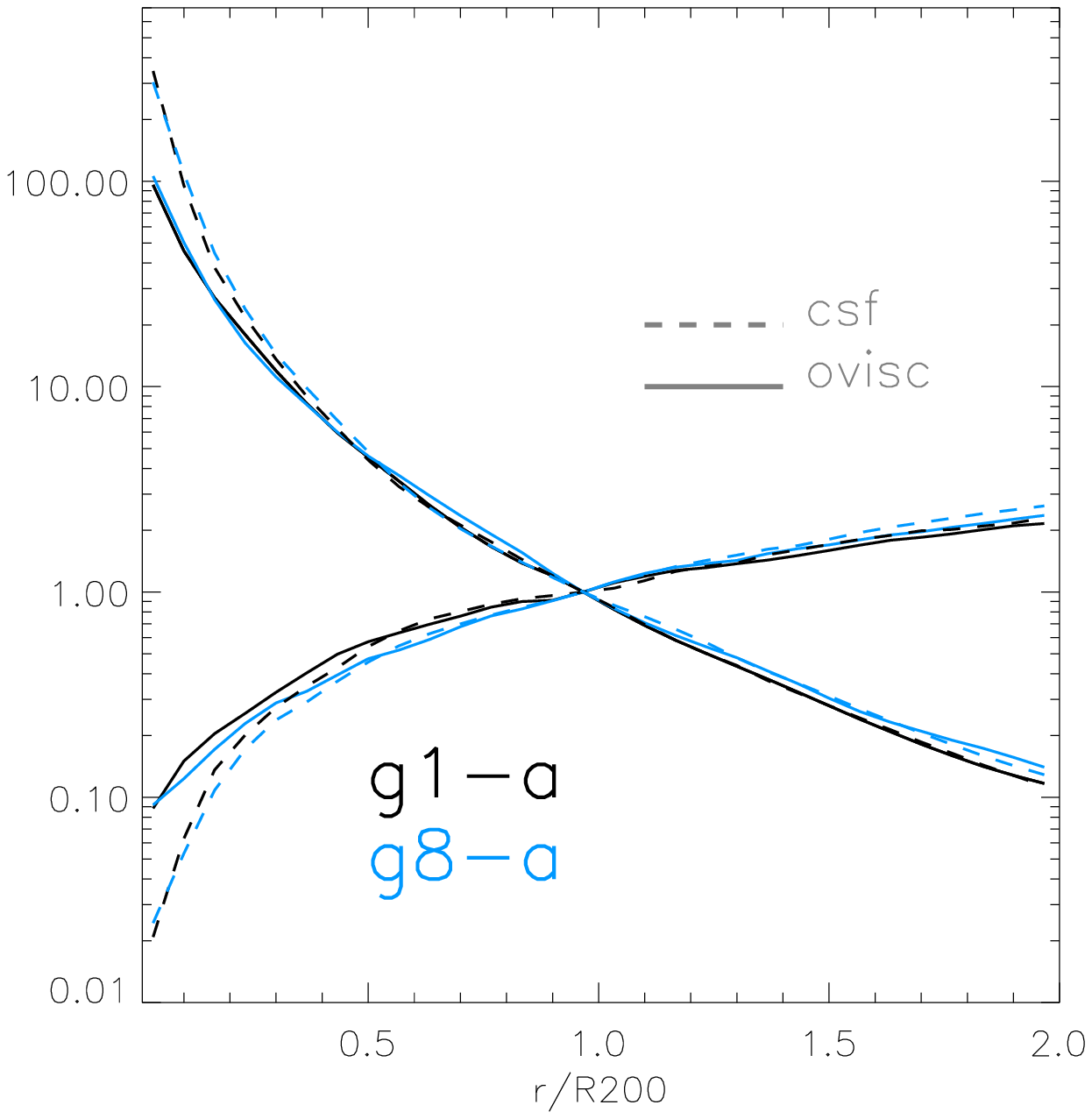}
\includegraphics[width=0.235\textwidth,height=0.23\textwidth]{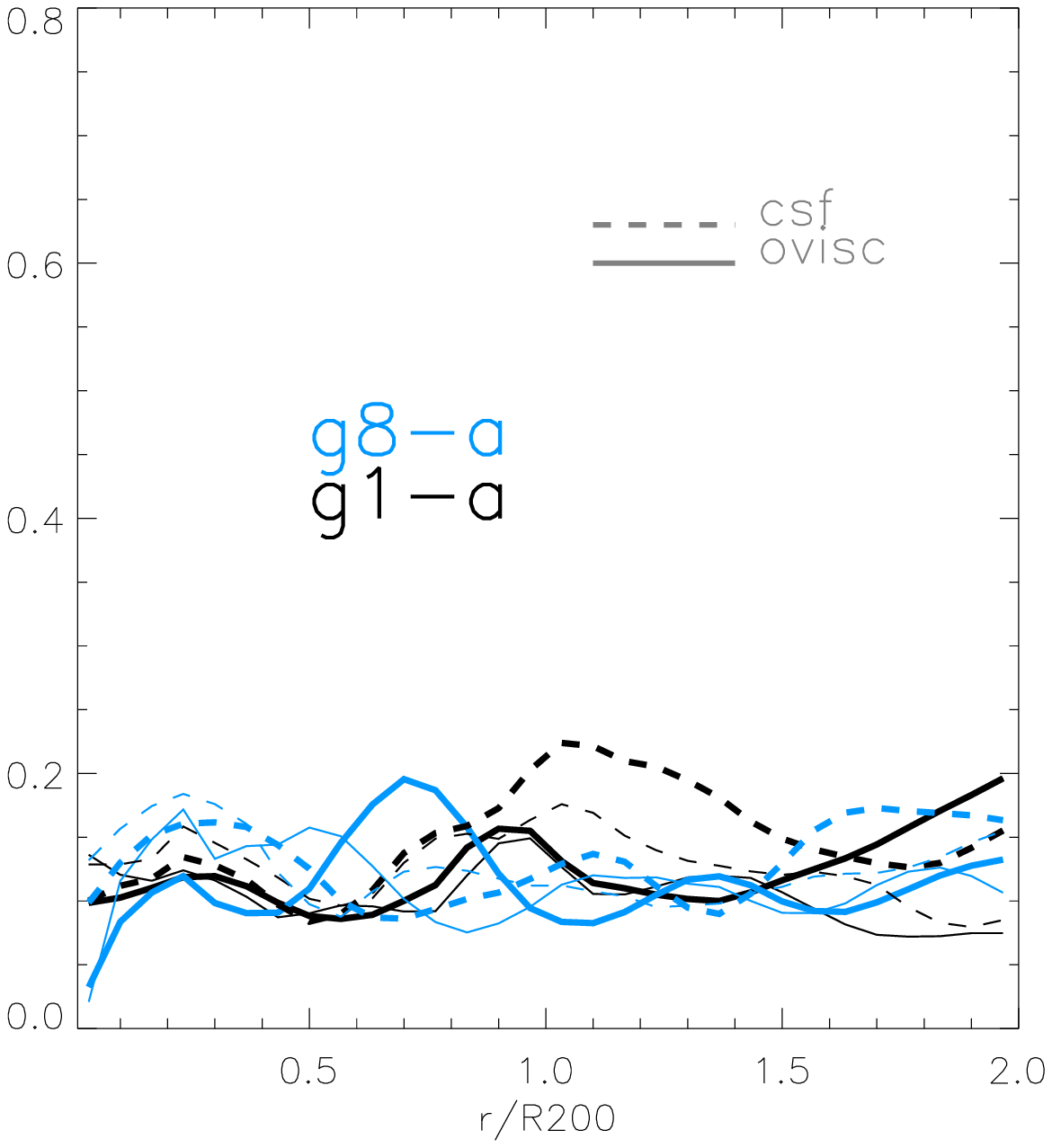}
\caption{{\it Left} panel: radial profiles for gas density (top lines) and gas entropy (bottom lines) for
clusters g1a and g8a of the GADGET2 sample. The solid lines refer to non-radiative runs, while the 
dashed lines refer to runs with radiative cooling, star formation and feedback from galactic winds.
{\it Right} panel: profile of azimuthal scatter for $N=8$ sectors. The thick lines refer to gas density,
the thin lines refer to gas entropy; the meaning of colors and line-styles is as in the Left panel.}
\label{fig:appendix}
\end{center}
\end{figure}

First, we show the  additional results for the profiles and the azimuthal scatter for the
the two most massive clusters of the GADGET2 runs (g1a and g8a), in re-simulations where cooling, 
star formation and feedback from galactic winds where self-consistently followed during the run.
The star formation is followed by adopting a sub-resolution multiphase model for the interstellar medium, including
also the feedback from supernovae and galactic outflows (Springel \& Hernquist 2003). In these runs, the efficiency
of supernovae to power galactic winds has been set to 50 per cent, which turns into a wind speed of $\sim 340km/s$. These re-simulations have been extracted from the same "{\it cfs}" dataset of Roncarelli et al.(2006). 
In the top panel of Fig.\ref{fig:appendix} we show the average gas density (top lines) and gas entropy (bottom lines) profiles for g1a and g8a at z=0, after the "99 percent" filtering procedure (Sec.\ref{subsec:method}), for
the reference physical run adopted in the main article (solid lines) and for the {\it csf} model (dashed).
Except for the sizable increase of the core gas density  and the decrease of the core gas entropy, as an obvious 
effect of radiative cooling, the agreement for all the radii outside $r>0.2 R_{\rm 200}$ is remarkable.
This is in line with the early results already reported by Roncarelli et al.(2006) and Romeo et al.(2006).

In the bottom panel of Fig.\ref{fig:appendix} we show the corresponding radial trend of the azimuthal scatter as defined in Eq.\ref{eq:scatter} and keeping $N=8$ sectors (the bold lines refer to gas density and the thin lines refer to gas entropy).
For both clusters in this case the differences in the azimuthal scatter are sizable when the two physical runs are compared ($\sim 10$ per cent at all radii); however all profiles scatter around the value of $S(r) \sim 0.10-0.15$ at all radii from $r>0.5R_{\rm 200}$, with no evident trend with radius. 
This may be well explained by the small scale effect of cooling sub-clumps at all radii in the {\it csf} run: the "99 per cent filtering" procedure outlined in the main paper efficiently removes the pointlike contribution from these clumps, however cooling make them more compact and enhances their survival within the cluster atmosphere and their efficiency in injecting shocks and small scale disturbances in the ICM (e.g. Valdarnini 2010), which can slightly change the local scatter behaviour, even if not the overall trend of azimuthal scatter in the clusters.
These results make us confident that the adoption of more complex physical modeling compared to the
non-radiative runs investigated in the main paper cannot change our conclusion in any sizable way, for all
the cluster volume outside of cores. 

\begin{figure}
\begin{center}
\includegraphics[width=0.36\textwidth,height=0.33\textwidth]{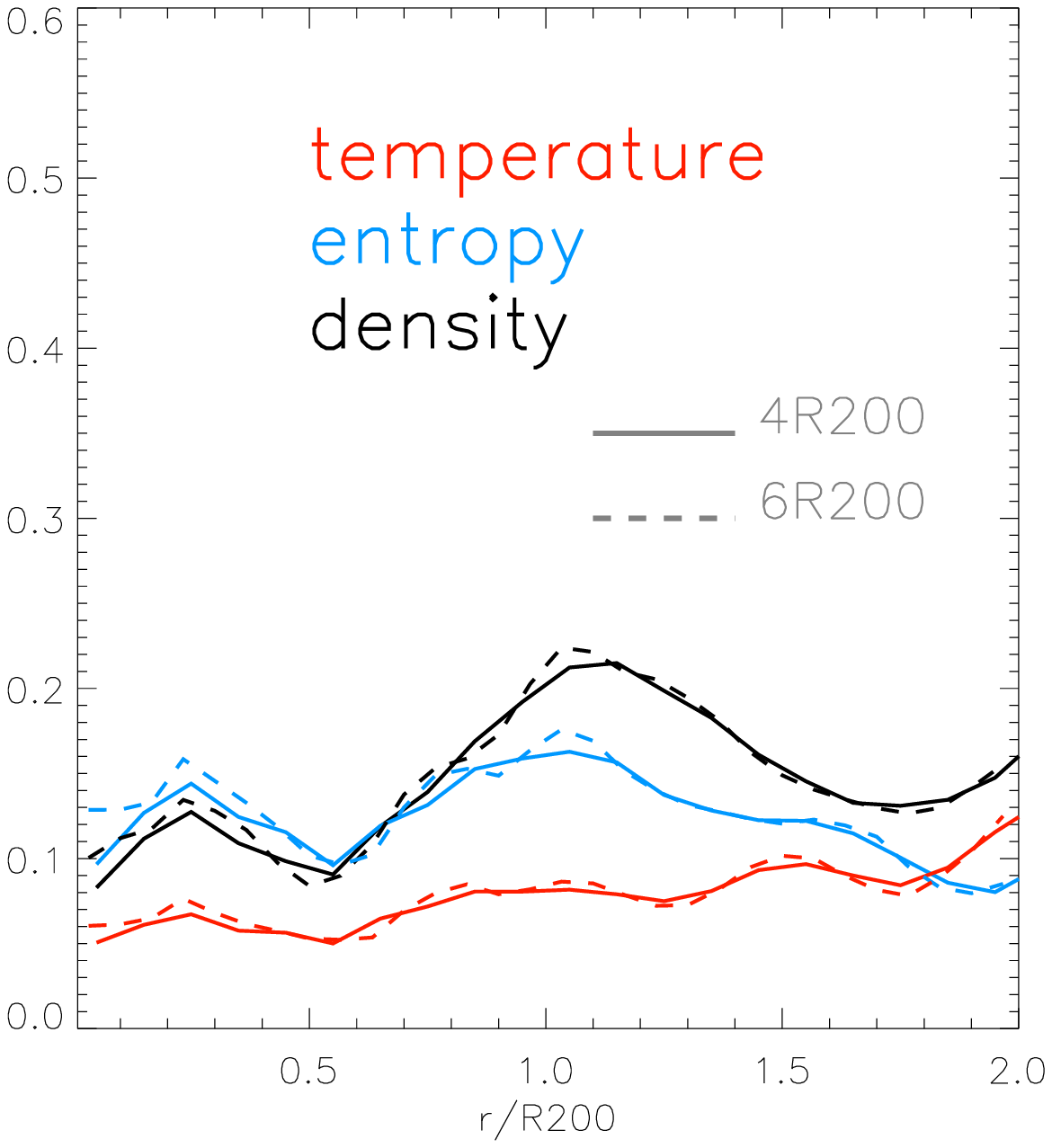}
\caption{Radial profile of azimuthal scatter of gas density, gas temperature and gas entropy for cluster g1a ($N=8$ sectors). The solid lines show the
results for a line of sight of $4 \cdot R_{200}$, the dashed lines show the results for a line of sight of $6 \cdot R_{200}$.}
\label{fig:appendix2}
\end{center}
\end{figure}

In a second test, we checked the stability of our results against the adoption of a larger line of sight across the projected cylinder, increasing its size from $4 \cdot R_{200}$ (as used in the main paper) to $6 \cdot R_{200} $. This was motivated by the fact that the overlap of not virialized regions and large scale environment along the line of sight towards the cluster may vary the reported values of the azimuthal scatter.
We thus re-computed the azimuthal scatter of gas density, gas temperature and gas entropy for cluster g1a of the GADGET2 sample, by considering a line
of sight of $6 \cdot R_{200}$  (Fig.\ref{fig:appendix2}). 
The test clearly shows that the difference in the azimuthal scatter of all quantities remains essentially the same within a few percent, with all trends un-altered. We thus conclude that our results are well constrained for a total line of sight of $\geq 4 \cdot R_{200}$.

\begin{thebibliography} {200}

\bibitem[Bautz et al.(2009)]{2009PASJ...61.1117B} Bautz, M.~W., et al.\ 
2009, PASJ, 61, 1117 

\bibitem[Burns et al.(2010)]{2010ApJ...721.1105B} Burns, J.~O., Skillman, 
S.~W., \& O'Shea, B.~W.\ 2010, ApJ, 721, 1105 



\bibitem[Cassano et al.(2010)]{2010ApJ...721L..82C} Cassano, R., Ettori, 
S., Giacintucci, S., Brunetti, G., Markevitch, M., Venturi, T., 
\& Gitti, M.\ 2010, ApJL, 721, L82 

\bibitem[Cavaliere et al.(2010)]{2010arXiv1010.4415C} Cavaliere, A., Lapi, 
A., \& Fusco-Femiano, R.\ 2010, arXiv:1010.4415 


\bibitem[\protect\citeauthoryear{Dolag et al.}{2005}]{2005MNRAS.364..753D} 
Dolag K., Vazza F., Brunetti G., Tormen G., 2005, MNRAS, 364, 753 

\bibitem[Ettori et al.(1998)]{1998MNRAS.300..837E} Ettori, S., Fabian, 
A.~C., \& White, D.~A.\ 1998, MNRAS, 300, 837 

\bibitem[Ettori 
\& Balestra(2009)]{2009A&A...496..343E} Ettori, S., \& Balestra, I.\ 2009, A\&A, 496, 343 


\bibitem[Ettori 
\& Molendi(2010)]{2010arXiv1005.0382E} Ettori, S., \& Molendi, S.\ 2010, arXiv:1005.0382 

\bibitem[Frenk et al.(1999)]{1999ApJ...525..554F} Frenk, C.~S., et al. 
1999, ApJ , 525, 554 

\bibitem[George et al.(2009)]{2009MNRAS.395..657G} George, M.~R., Fabian, 
A.~C., Sanders, J.~S., Young, A.~J., 
\& Russell, H.~R.\ 2009, MNRAS, 395, 657 

\bibitem[Hoshino et al.(2010)]{2010arXiv1001.5133H} Hoshino, A., et al.\ 
2010, arXiv:1001.5133 

\bibitem[Kawahara et al.(2008)]{2008ApJ...687..936K} Kawahara, H., Reese, 
E.~D., Kitayama, T., Sasaki, S., \& Suto, Y.\ 2008, ApJ, 687, 936 

\bibitem[Kawaharada et al.(2010)]{2010ApJ...714..423K} Kawaharada, M., et 
al.\ 2010, ApJ, 714, 423 


\bibitem[Lapi et 
al.(2010)]{2010A&A...516A..34L} Lapi, A., Fusco-Femiano, R., \& Cavaliere, A.\ 2010, A\&A, 516, A34 


\bibitem[Leccardi 
\& Molendi(2008)]{2008A&A...486..359L} Leccardi, A., \& Molendi, S.\ 2008, A\&A, 486, 359 

\bibitem[Liedahl et al.(1995)]{1995ApJ...438L.115L} Liedahl, D.~A., 
Osterheld, A.~L., \& Goldstein, W.~H.\ 1995, ApJL, 438, L115 

\bibitem[Loken et al.(2002)]{2002ApJ...579..571L} Loken, C., Norman, M.~L., 
Nelson, E., Burns, J., Bryan, G.~L., \& Motl, P.\ 2002, ApJ, 579, 571 


\bibitem[Mitchell et al.(2009)]{2009MNRAS.395..180M} Mitchell, N.~L., 
McCarthy, I.~G., Bower, R.~G., Theuns, T., 
\& Crain, R.~A. 2009, MNRAS, 395, 180 

\bibitem[Neumann(2005)]{2005A&A...439..465N} Neumann, D.~M.\ 2005,A\&A, 439, 465 

\bibitem[\protect\citeauthoryear{Norman et al.}{2007}]{2007arXiv0705.1556N} 
Norman M.~L., Bryan G.~L., Harkness R., Bordner J., Reynolds D., O'Shea B., 
Wagner R., 2007, arXiv, 705, arXiv:0705.1556 

\bibitem[Rasia et al.(2005)]{2005ApJ...618L...1R} Rasia, E., Mazzotta, P., 
Borgani, S., Moscardini, L., Dolag, K., Tormen, G., Diaferio, A., 
\& Murante, G.\ 2005, ApJL, 618, L1 

\bibitem[Romeo et al.(2006)]{2006MNRAS.371..548R} Romeo, A.~D., 
Sommer-Larsen, J., Portinari, L., 
\& Antonuccio-Delogu, V.\ 2006, MNRAS, 371, 548 

\bibitem[Roncarelli et al.(2006)]{2006MNRAS.373.1339R} Roncarelli, M., 
Ettori, S., Dolag, K., Moscardini, L., Borgani, S., 
\& Murante, G.\ 2006, MNRAS, 373, 1339 

\bibitem[Reiprich et 
al.(2009)]{2009A&A...501..899R} Reiprich, T.~H., et al.\ 2009, A\&A, 501, 899 

\bibitem[Springel 
\& Hernquist(2003)]{2003MNRAS.339..289S} Springel, V., \& Hernquist, L.\ 2003, MNRAS, 339, 289 


\bibitem[\protect\citeauthoryear{Springel}{2005}]{2005MNRAS.364.1105S}
Springel V., 2005, MNRAS, 364, 1105

\bibitem[\protect\citeauthoryear{Springel}{2010}]{2010MNRAS.401..791S} 
Springel V., 2010, MNRAS, 401, 791 

\bibitem[Valdarnini(2010)]{2010arXiv1010.3378V} Valdarnini, R.\ 2010, 
arXiv:1010.3378 

\bibitem[Vazza et al.(2010)]{2010NewA...15..695V} Vazza, F., Brunetti, G., 
Gheller, C., \& Brunino, R.\ 2010, NewA, 15, 695 

\bibitem[Vazza(2010)]{2010MNRAS.tmp.1724V} Vazza, F.\ 2010, MNRAS, 1724 

\bibitem[Voit et al.(2005)]{2005MNRAS.364..909V} Voit, G.~M., Kay, S.~T., 
\& Bryan, G.~L. 2005, MNRAS, 364, 909 

\bibitem[Vikhlinin et al.(1999)]{1999ApJ...525...47V} Vikhlinin, A., 
Forman, W., \& Jones, C.\ 1999, ApJ, 525, 47 

\bibitem[Vikhlinin et al.(2006)]{2006ApJ...640..691V} Vikhlinin, A., 
Kravtsov, A., Forman, W., Jones, C., Markevitch, M., Murray, S.~S., 
\& Van Speybroeck, L.\ 2006, ApJ, 640, 691 

\bibitem[Wadsley et al.(2008)]{2008MNRAS.387..427W} Wadsley, J.~W., 
Veeravalli, G., \& Couchman, H.~M.~P. 2008, MNRAS, 387, 427 




  \end{thebibliography}
 \end{document}